\documentclass[aps,pra,twocolumn,groupedaddress]{revtex4-1}

\usepackage{amsmath}
\usepackage{amsthm}
\usepackage{amssymb}
\usepackage{graphicx,hyperref,bbm,enumitem,bm}  
\usepackage[pdftex]{color}

\usepackage[toc,page]{appendix}

\def\ii{{\rm i}}
\def\sx{\sigma^{\rm x}}
\def\sy{\sigma^{\rm y}}
\def\sz{\sigma^{\rm z}}
\def\tr#1{{\rm tr}{#1}}
\def\1{\mathbbm{1}}
\def\ket#1{{| #1 \rangle}}
\def\bra#1{{\langle #1 |}}
\def\braket#1#2{{\langle #1 | #2 \rangle}}

\def\ax{a_{\rm x}}
\def\ay{a_{\rm y}}
\def\az{a_{\rm z}}

\def\n2{{\lfloor \frac{n}{2} \rfloor}}

\def\nA{{n_{\rm A}}}
\def\nB{{n_{\rm B}}}

\def\IA{I_{\rm A}}
\def\s{\boldsymbol{ \sigma }}
\def\vLR{v_{\mathrm{LR}}}
\def\vB{v_{\mathrm{B}}}

\def\tit#1{{\em #1},}

\usepackage{booktabs}
\usepackage{multirow}
\AtBeginDocument{
\heavyrulewidth=.08em
\lightrulewidth=.05em
\cmidrulewidth=.03em
\belowrulesep=.65ex
\belowbottomsep=0pt
\aboverulesep=.4ex
\abovetopsep=0pt
\cmidrulesep=\doublerulesep
\cmidrulekern=.5em
\defaultaddspace=.5em
}

\begin{document}
	
	\title{Two-step phantom relaxation of out-of-time-ordered correlations in random circuits}
	
	\author{Ja\v s Bensa and Marko \v Znidari\v c}
	\affiliation{Department of Physics, Faculty of Mathematics and Physics, University of Ljubljana, 1000 Ljubljana, Slovenia}
	
	\date{\today}
	
	\begin{abstract}
We study out-of-time-ordered correlation (OTOC) functions in various random quantum circuits and show that the average dynamics is governed by a Markovian propagator. This is then used to study relaxation of OTOC to its long-time average value in circuits with random single-qubit unitaries, finding that relaxation in general proceeds in two steps: in the first phase that lasts upto an extensively long time the relaxation rate is given by a phantom eigenvalue of a non-symmetric propagator, whereas in the second phase the rate is determined by the true 2nd largest propagator eigenvalue. We also obtain exact OTOC dynamics on the light-cone and an expression for the average OTOC in finite random circuits with random two-qubit gates.
	\end{abstract}
	
	\maketitle
	
	\section{Introduction}

        Complexity of quantum evolution is of wide theoretical and practical interest. It can be captured in different ways, one common idea is to quantify what is colloquially called scrambling of quantum information~\cite{Hayden07,Susskind08}. Details of what scrambling means depend on a particular situation, but broadly speaking one can quantify it in two ways: either by properties of the evolved state, for instance its entanglement under random circuit evolution~\cite{emerson03}, or by complexity of time-evolved operators as measured for instance by the out-of-time-ordered correlation (OTOC) functions~\cite{lashkari13,shenker14,maldacena16}. In the present paper we shall study dynamics of OTOC $O^\beta(i,j,t)$, being equal to the Hilbert-Schmidt norm of the commutator, $O^\beta(i,j,t)=\frac{1}{2}\langle AA^\dagger\rangle=\frac{1}{2^{n+1}}\tr{(AA^\dagger)}$, where $A=[\sigma_i^\alpha(t), \sigma_j^\beta]$ is a commutator between two local traceless operators, one of them being evolved in time. It therefore measures how fast correlations spread from the spatial position $i$ to $j$, and, more importantly for our discussion, also how fast $\sigma_i^\alpha(t)$ becomes ``random''.

        Due to its relative simplicity and relevance for quantum information OTOCs have been studied in very many different contexts. Limiting just to homogeneous systems, these include field theory~\cite{roberts15,swingle17}, Luttinger liquids~\cite{moessner17}, and (chaotic) many-body systems~\cite{knap17,ueda17,xu19,lin18,nakamura19,smith19}. In studies of quantum many-body systems one has to usually resort to numerics and that is why any exact results are greatly appreciated. Simplification that allows for analytical results often comes due to symmetries, either exact ones (e.g., integrable systems) or for instance effectively increasing symmetry by some averaging procedure. One such case when averaging brings simplification is that of random circuits where it has been shown that the average dynamics can be described by a Markov chain~\cite{oliveira}, leading to exact entanglement generation speeds for specific circuits~\cite{PRA08}. Today many other theoretical approaches to random circuits are known. A prominent example is generic hydrodynamic description of operator spreading and OTOC dynamics~\cite{Frank18,adam18}, explicitly verified by exact results for random U(4) circuits. For a slightly different Brownian Hamiltonian evolution see Ref.~\cite{zhou19}. Another powerful example of exactly solvable dynamics are so called dual-unitary circuits~\cite{DU-LC}, among them also random dual-unitary circuits~\cite{Bruno20}. One of the distinguished features of dual-unitary circuits is that 2-point spatio-temporal correlations are nonzero only on the light-cone boundary~\cite{DU-LC}, and one can use a powerful finite transfer matrix formalism. Similar is the case in integrable circuits~\cite{lamacraft21} in which the gates satisfy the Yang-Baxter equation. For dual-unitary circuits OTOC decay exponentially with time close to the light-cone boundary~\cite{maximum_velocity}. Some simplification is possible also for certain small perturbations to dual-unitarity~\cite{kos21}.

Many studies try to find some indication of chaoticity. Remembering that the notion of chaos is in classical systems defined as a property of the long-time limit, one might ask if such long-time complexity is somehow reflected also in the OTOC dynamics. The answer is not clear, what one can say however is that for lattice systems with finite local Hilbert space dimension it is not clear how to distinguish chaoticity from integrability via OTOCs. A possible approach is to get some measure of instability, like ``quantum'' Lyapunov exponents, from OTOCs dynamics. However, this is bound to fail for several reasons. One is that one might get an exponential behavior that is unrelated to chaos, for instance simply due to unstable fixed points~\cite{cao20,hashimoto20,santos20}. Hydrodynamic behavior of the operator front might also look the same in chaotic and integrable systems~\cite{sarang18} (for free systems see Ref.~\cite{riddell21}). On top of it, in lattice models with finite local dimension, like chains of qubits, there is no obvious small parameter and so any possible exponential Lyapunov-like growth of OTOCs can hold only upto finite (short) times~\cite{saso17,khemani18}.

We are going to study OTOC dynamics in random quantum circuits, mostly in one-dimensional geometry and for qubits. In random circuits there is no dichotomy between integrability and chaos -- random circuits can be thought of as being models of chaotic systems -- and so we are not aiming at coming up with some chaoticity criterion. What we shall focus on is the long-time dynamics of OTOCs, specifically on how fast OTOC relaxes to its asymptotic value reached at long times that corresponds to a completely scrambled evolution. As we shall see, this will reveal interesting mathematical and physical properties. 

\begin{figure}[t]
\centerline{\includegraphics[width=1.9in,angle=-90]{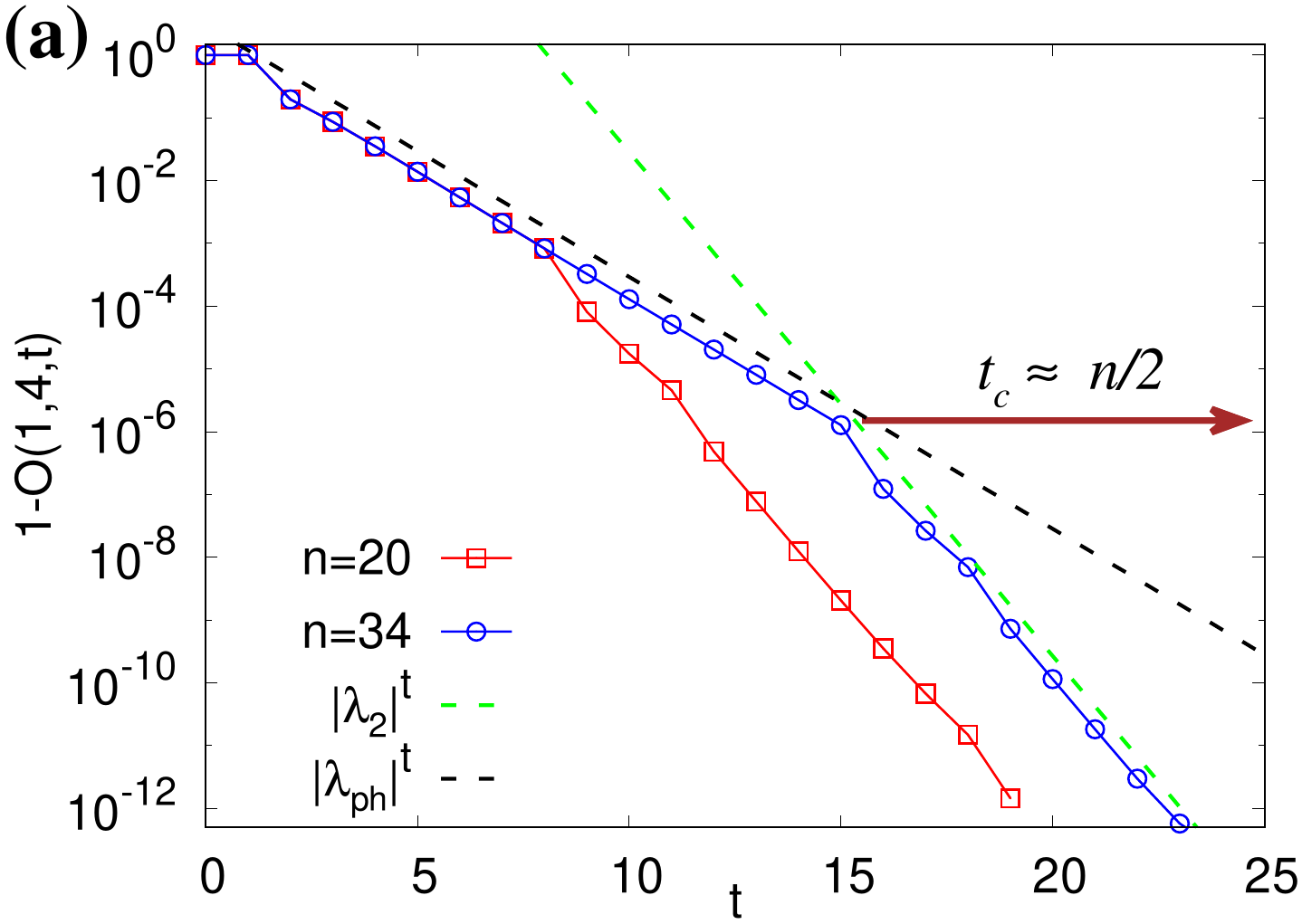}}
\vskip3mm
\centerline{\includegraphics[width=2.0in,angle=-90]{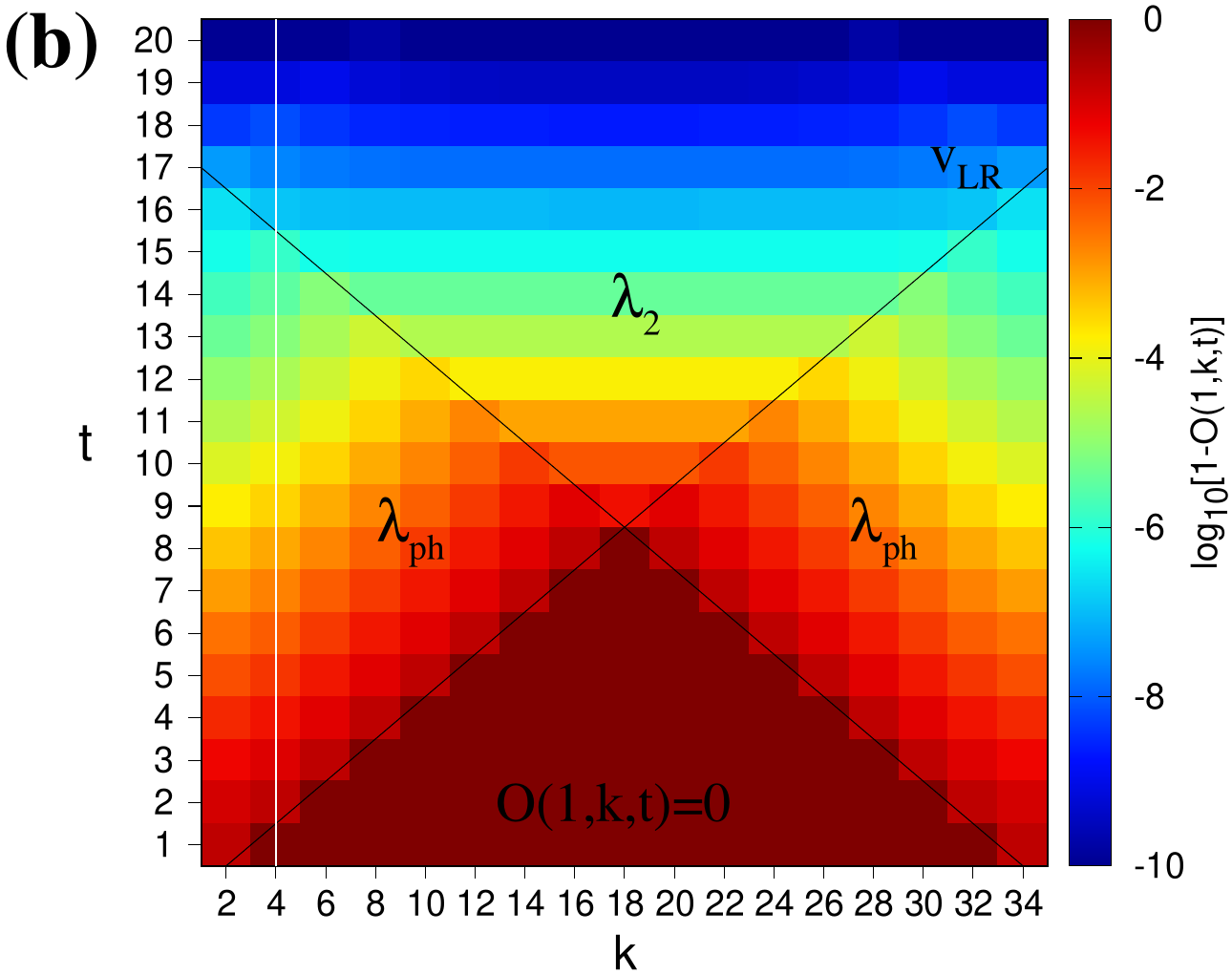}}
\caption{Phantom relaxation of OTOCs. (a) In the thermodynamic limit OTOC relaxes to its long-time value as $\lambda_{\rm ph}^t$, instead of $|\lambda_{2}|^t$, where $|\lambda_2|$ is the 2nd largest eigenvalue of the OTOC transfer matrix $M$, and $\lambda_{\rm ph}$ is a phantom eigenvalue (which is not an eigenvalue of $M$). (b) Spatio-temporal plot of OTOCs $O(1,k,t)$ for $n=34$ where one can see regions of relaxation with $\lambda_{\rm ph}$, and the asymptotic region with $\lambda_2$. White vertical line marks the cross-section shown in (a). All is for a random circuit with the XXZ gate with $\az=0.2$ and the brick-wall protocol with periodic boundary conditions.}
\label{fig:shema}
\end{figure}

Deriving a Markovian description of the average OTOC dynamics in random circuits we shall show that the relaxation rate typically exhibits a discontinuity at a specific time linear in the number of qubits. What is more, the relaxation time in this first phase, which is dominant in the thermodynamic limit, is not given by the gap of the Markovian matrix. Instead, it is given by a so-called phantom eigenvalue -- a fake ``eigenvalue'' that is not in the spectrum. Illustration of such phantom relaxation is in Fig.~\ref{fig:shema}. Looking at a particular OTOC $O(1,4,t)$ (see Eq.~(\ref{eq:OTOC_first_def}) and (\ref{eq:OTOC_final_def}) for definitions), whose dynamics is given by a particular Markovian matrix $M$, we study its relaxation towards $O(1,4,t\to \infty) \approx 1$. One can see that the relaxation proceeds in two steps: asymptotically at large $t> t_{\rm c}\approx n/2$ one has the expected exponential decay $\sim |\lambda_2|^t$, where $\lambda_2$ is the second largest eigenvalue of $M$; however, for $t<t_{\rm c}$ the exponential relaxation goes as $\sim \lambda_{\rm ph}^t$, where $\lambda_{\rm ph}$ is a ``phantom'' eigenvalue that is larger than any true eigenvalue of $M$. Because in the thermodynamic limit $t_{\rm c}$ diverges, the correct relaxation rate that one will observe at any finite time is not given by the spectrum of $M$, but instead by the phantom $\lambda_{\rm ph}$. Curiously, it turns out that in some cases $\lambda_{\rm ph}$ is equal to the 2nd largest eigenvalue of a different circuit not related in any obvious way to $M$.

Similar phenomenon of phantom relaxation has been recently observed also in purity dynamics~\cite{prejsnji_clanek}. Perhaps also related is an observation that in nonequilibrium dynamics described by the Lindblad equation the gap does not necessarily give the correct relaxation time~\cite{Mori20,ueda21}, and of non-Hermiticity of transfer matrix describing integrable circuits~\cite{lamacraft21}.

	\section{Random quantum circuits}

\begin{figure}[t]
		\begin{center}
	        \includegraphics[width=85mm]{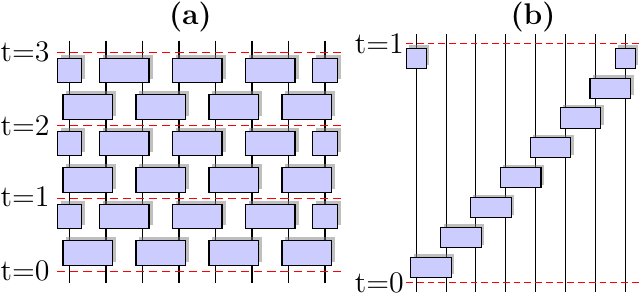}
			\caption{Illustration of a brick-wall (BW) protocol (a) and a staircase (S) protocol (b) on a qubit chain of size $n=8$ with periodic boundary conditions. Blue boxes represent elementary steps $U_{i,j}$. Red dotted lines represent integer times, which are measured so that one unit corresponds to the action of one period of the random quantum circuit. One period of a BW protocol consists of local operators $U_{i,i+1}$ where we first act on qubits with odd $i$, then on qubits with even $i$. In a S protocol in one period we subsequently act with elementary steps $U_{i,i+1}$ starting from $i=1$ and increasing $i$ by one for each local operator.}
			\label{fig:BWandS}
		\end{center}
	\end{figure}

	In this paper we deal with random quantum circuits defined on a system of $n$ qubits. The unitary propagator $U$ is a product of local elementary gates $U_{i,j}$ acting on qubit pairs $(i,j)$, that is $U = \prod_{i,j} U_{i,j}$. Every elementary step is, in turn, defined as a product of two independent one-site random unitaries $V_i$ and $V_j$ and a two-site unitary gate $W_{i,j}$; namely $U_{i,j} = W_{i,j} V_{i} V_{j}$. Two examples of random quantum circuits, where the product of elementary gates is ordered in a brick-wall (BW) pattern and in a staircase (S) pattern, can be seen in Fig.~\ref{fig:BWandS}. As can be deduced from the name, the BW protocol is defined as a configuration where in each unit of time we first couple nearest-neighbor qubits $(i,i+1)$ with an odd $i$, then all pairs with even $i$. Apart from being widely studied for its simplicity, we mainly focus on this protocol because it turned out to be the fastest possible local scrambler of entanglement \cite{prejsnji_clanek}. Another configuration that we will encounter in this paper is the S configuration. The S configuration consists of operators $U_{i,i+1}$, where at each step we increase $i$ by $1$. In the main part we shall mostly focus on random quantum circuits acting on 1-dimensional (1D) chains of qubits with either open boundary conditions (OBC) or periodic boundary conditions (PBC); that is, qubits are distributed on a line (OBC) or on a circle (PBC).

One obtains various random circuits by different choices of a fixed two-site gate $W_{i,j}$ and the ordering of elementary steps. To distinguish various choices of $W_{i,j}$ we shall parametrize it in the following canonical form \cite{dekompozicija_1,dekompozicija_2,dekompozicija_recept}
	
	\begin{align}
		W_{j,k} &= V_{j}^{(1)} V_{k}^{(2)} w_{j,k} (\textbf{a})V_{j}^{(2)} V_{k}^{(3)} \nonumber \\
		w_{j,k}(\textbf{a}) &= \exp \left[ \ii \frac{\pi}{4} \left( \ax \sx_j \sx_k + \ay \sy_j \sy_k + \az \sz_j \sz_k \right)   \right],
		\label{eq:parametrizacija}
	\end{align}
	where $V_k^\alpha$ are one-site unitary operators, $\sigma^{\rm x,\rm y,\rm z}$ are Pauli matrices and $\textbf{a} = (\ax, \ay , \az)$ are three real parameters, which can be constrained to $0 \leq \az \leq \ay \leq \ax \leq 1$. In this paper, we will be interested in the average dynamics of OTOCs generated by random quantum circuits. Due to randomness on single qubits (at every elementary step we act with random unitaries $V_{i}$ and $V_{j}$) the choice of local operators $V_k^\alpha$ does not affect our averaged dynamics, so only the choice of the three real parameters $(\ax, \ay , \az)$ is what matters. To conclude, without loss of generality we can take our fixed two-site unitary to be $w_{i,j}$, which is in turn parametrized by only three constrained real parameters $0 \leq \az \leq \ay \leq \ax \leq 1$. 
		
\section{OTOC Markov chain}
	
We shall study out-of-time-order correlations defined as
	\begin{align}
		O^\beta(i,j,t) &= \frac{1}{2^{n+1}} \tr{\,\vline\, \left[ \sigma_i^\alpha(t), \sigma_j^\beta \right]\space\vline\,}^2 \nonumber \\
		&= 1-\frac{1}{2^n}\tr{\left( \sigma_i^\alpha(t) \sigma_j^\beta \sigma_i^\alpha(t) \sigma_j^\beta \right)},
		\label{eq:OTOC_first_def}
	\end{align} 
	with $\sigma_k^\gamma$ denoting the Pauli matrix at position $k$, and $\gamma \in \{\rm x,\rm y,\rm z\}$. The time-evolved Pauli matrix is obtained as $\sigma_i^\alpha(t)=U^\dagger \sigma_i^\alpha U$. OTOC thus measures how correlations between two initially localized operators spread in the system. Its minimal value is $0$ for operators that commute, i.e., until $\sigma_i^\alpha(t)$ begins to overlap with $\sigma_j^\beta$, whereas its maximal value is $2$ reached for e.g., $\sigma_j^\beta=\sx_j$ and $\sigma_i^\alpha(t)=\sy_j$. If $\sigma_i^\alpha(t)$ at large times randomly spreads over all available operator space the average OTOC will converge towards its thermal value $O_\infty \approx 1$ (see Appendix~\ref{app:O_inf}). We are going to study how OTOCs converge to this long-time stationary value. Note that often the name OTOC is used just for the 2nd term in Eq.~(\ref{eq:OTOC_first_def}), whose asymptotic value goes towards $0$. Because we will be interested in relaxation we will in fact study this 2nd term.

	It has been shown that averaging over one-site random unitaries leads to a Markov chain description of the evolution of the average purity~\cite{oliveira07,metoda_redukcija}. Because OTOCs are, similarly as purity, also quadratic in the time-evolved operator, their average evolution can also be written in terms of a Markov chain. This has been done for the special case of a random U(4) elementary step $U_{i,j}$ in Ref.~\cite{adam18}, whereas we derive the Markovian matrix description for a protocol consisting of an arbitrary two-qubit $W_{i,j}$ conjugated by independent single-qubit unitaries. 

The derivation relies on the fact that it is possible to express OTOCs as a linear combination of all possible purities of a system of $n$ qubits. Writing the operator $\sigma_i^\alpha(t)$ in the basis of Pauli strings with coefficients $a_{\s}(t)$
	\begin{equation}
		\sigma_i^\alpha(t) = \sum_{\s} a_{\s}(t)\, \vec{\s}, 
\label{eq:salpha}
	\end{equation}
	 where we use the label $\s = (\sigma_1,\sigma_2,\dots,\sigma_n)$ while the product of Pauli matrices on all sites is denoted by $\vec{\s}=\sigma_1 \sigma_2\cdots \sigma_n$, and $\sigma_i \in \{ \1,\sx,\sy,\sz \}_i$, we obtain
	\begin{equation}
		O^\beta(i,j,t) = 2 \sum_{\s; \sigma_j \in S_1 \setminus \sigma_j^\beta} a_{\s}^2(t),
		\label{eq:Obeta}
	\end{equation}
where for brevity we defined two sets
\begin{equation}
S_0=\{\1 \},\qquad  S_1=\{ \sigma^x,\sigma^y,\sigma^z\}
\label{eq:S}
\end{equation}
that will be useful in specifying various summations. For instance the sum in Eq.~(\ref{eq:Obeta}) runs over all Pauli strings $\s = (\sigma_1,\sigma_2,\dots,\sigma_n)$ except for those having $\sigma_j^\beta$ or $\1$ at the site $j$.
	
	We wish to relate a vector containing all possible OTOCs $O(i,j,t)$ for every position $j$ to a vector of purities through a linear transformation. To obtain purity we write the density operator in terms of Pauli strings coefficients $c_{\s}$, $\rho(t)=\frac{1}{\sqrt{2^n}}\sum_{\s}{c_{\s}\, \vec{\s}}$. Purity $\IA$, which measures pure-state entanglement between two complementary subsets of qubits denoted by $\rm A$ and $\rm B$ (consisting of $\nA$ and $\nB$ qubits, respectively), is then
	
	\begin{equation}
		\IA = {\rm tr}_{\rm A} \left( {\rm tr}_{\rm B} \rho \right)^2 = 2^{\nB} \sum_{\s; \forall i \in \rm B, \sigma_i=\1} c_{\s}^2.
		\label{eq:IA}
	\end{equation}	
Expression (\ref{eq:IA}) is invariant with respect to an arbitrary permutation of the three Pauli matrices at any site. In other words, it is only the totally symmetric sum of $c_{\s}^2$ for all three Pauli matrices that matters for purity. For instance, for a system of two qubits with subsystem $\rm A$ being the 1st qubit, we have $\IA=2 c^2_{(\1,\1)}+2(c^2_{(\sigma^{\rm x},\1)}+c^2_{(\sigma^{\rm y},\1)}+c^2_{(\sigma^{\rm z},\1)})$. So instead of bookkeeping all $4^2$ coefficients $c^2_{(\sigma_1.\sigma_2)}$ it is enough to keep track of only $2^2$ combinations of them, which we can neatly pack into a two-site vector (for definition of $S_1$ see Eq.(\ref{eq:S}))
	\begin{equation}
		\Phi = \begin{pmatrix}
			c_{(\1,\1)}^2	\\
			\sum_{\sigma_1 \in S_1} c_{(\sigma_1,\1)}^2	\\
			\sum_{\sigma_2 \in S_1} c_{(\1,~\sigma_2)}^2	\\
			\sum_{\sigma_1,\sigma_2 \in S_1} c_{(\sigma_1,\sigma_2)}^2
		\end{pmatrix}.
	\label{eq:In2}
	\end{equation}
We can obtain purities for all possible bipartitions of two qubits from components of $\Phi$, specifically, if the 1st qubit is in $\rm A$ we have $\IA=2\Phi_{0}+2\Phi_{1}$, whereas if the 2nd qubit is in $\rm A$ one has $\IA=2\Phi_{0}+2\Phi_2$, where we labeled the 4 components in Eq.~(\ref{eq:In2}) by $\Phi_{0,1,2,3}$. Generalizing $\Phi$ to $n$ qubits it will have $2^n$ components that we label by bit strings $\mathbf{s}=(s_1,\ldots,s_n)$, where $s_j\in \{0,1\}$, with the components being 
\begin{equation}
\Phi_{\mathbf{s}}=\sum_{\s;\sigma_j \in S_{s_j}} c^2_{\s}.
\label{eq:Phis}
\end{equation}
To shorten the notation we shall occasionally also use the integer value of the bit string $\mathbf{s}$ instead of specifying the full $\mathbf{s}=(s_1,\ldots,s_n)$, as $\mathbf{s} \equiv \sum_{j=1}^n 2^{j-1} s_j$. Purity for an arbitrary bipartition is now given by a particular component of vector $\Phi_I$ obtained as
	\begin{equation}
		\Phi_I := A_I \Phi, \quad
		A_I = \begin{pmatrix}
			1 & 1	\\
			2 & 0
		\end{pmatrix}^{\otimes n},
		\label{eq:I_sum}
	\end{equation}
Specifically, the component $[\Phi_I]_{\mathbf{s}}$ is equal to the purity for a bipartition in which the subsystem A consists of qubits for which $s_j=0$, i.e., the bit $s_j$ encodes the subsystem in which the $j$-th qubit is.
	
	Ref.~\cite{metoda_redukcija} showed that it is possible to write the evolution of purities $\Phi_I$ averaged over one-site Haar random unitaries as a Markov chain. Abusing notation and from now on using $\Phi_I(t)$ to denote the average purity after $t$ steps of our random circuits (\ref{fig:BWandS}), one has
	\begin{equation}
		\Phi_I(t) = M' \Phi_I(t-1).
\label{eq:Phit}
	\end{equation}
The transfer matrix $M'$ describing one period of our circuit is a product of matrices $M'_{i,j}$, one for each elementary step $U_{i,j}$~\cite{metoda_redukcija}. For example, a transfer matrix describing $t_2$ periods of a BW PBC circuit on $n=4$ qubits would be $(M')^{t_2} = (M'_{4,1} M'_{2,3} M'_{3,4} M'_{1,2})^{t_2}$. Note that because the two-site gates $W_{i,j}$ are the same for all steps all transfer matrices are independent of time. 
	
Looking at the expressions for OTOC in Eq.~(\ref{eq:Obeta}) and $\Phi$ in Eq.~(\ref{eq:Phis}) we can see that they look rather similar. Because we know how average purities are evolved (\ref{eq:Phit}), we also know how to evolve $\Phi(t)$, namely defining $\Phi(t)=A_I^{-1} \Phi_I(t)$ gives us $\Phi(t)=A_I^{-1}M'A_I\Phi(t-1)$. This will in turn lead us to the evolution of OTOC. 

To achieve that let us rather look at the OTOC averaged over three possible $\sigma_j^\beta$,
\begin{equation}
		O(i,j,t) := \frac{1}{3} \sum_{\beta \in \{ \rm x,\rm y, \rm z \}} O^\beta(i,j,t) = \frac{4}{3} \sum_{\s; \sigma_j \in S_1} a_{\s}^2(t).
		\label{eq:OTOC_final_def}
	\end{equation}
Note that the dependence on site index $i$ is implicitly hidden in the expansion coefficients $a_{\s}(t)$ (\ref{eq:salpha}) of the initial $\sigma_i^\alpha$. Using $\Phi$ for a vector defined as in Eq.~(\ref{eq:Phis}) but for coefficients $a_{\s}$, and formally defining a vector $\Phi_O$ by
	\begin{equation}
		\Phi_O := A_O \Phi, \quad
		A_O = \begin{pmatrix}
			1 & 1	\\
			0 & \frac{4}{3}
		\end{pmatrix}^{\otimes n},
		\label{eq:O_sum}
	\end{equation}
one can verify that $O(i,j,t)$ is equal to the $2^{j-1}$-th component of the vector $\Phi_O$. That is, $O(i,j,t)=[\Phi_O]_{\mathbf{s}}$, where $s_j=1$ and $s_{k\neq j}=0$. Therefore, $n$ components of $\Phi_O$ are equal to OTOCs while the other $2^n-n$ components are some other combinations of $a^2_\mathbf{s}$ not related to OTOCs. Note that the choice of $A_O$ is not unique; the two $1$ in the top row take care of summing over both sets $S_0$ and $S_1$ for sites $k\neq j$ in Eq.~(\ref{eq:OTOC_final_def}), while the $\frac{4}{3}$ in the 2nd row accounts for an overall prefactor accounted by a single bit $s_j$ being $1$, ie., summation only over $S_1$ at site $j$. The initial value of OTOC $O(i,j,0)$ is easily computed from the initial value of $a_{\s}=\delta_{\s,(\1,\ldots,\1,\sigma_i^\alpha,\1,\ldots,\1)}$ ($\delta_{\s,\s'}=\Pi_k \delta_{\sigma_k,\sigma'_k}$ is a Kronecker multi-delta), which in turn gives (\ref{eq:Phis}) $[\Phi(t=0)]_{\mathbf{s}}=\delta_{\mathbf{s},(0,\ldots,0,1_i,0,\ldots,0)}=\delta_{\mathbf{s},2^{i-1}}$, which then through (\ref{eq:O_sum}) results in the initial condition
	\begin{equation}
		\Phi_O(t=0) = \frac{4}{3} \textbf{e}_{2^{i-1}}+\textbf{e}_{0},
		\label{eq:OTOC_initial}
	\end{equation}
	where the vector $\textbf{e}_{k}$ has components $[\textbf{e}_{k}]_{\mathbf{s}}=\delta_{\mathbf{s},k}$. The vector $\Phi$ containing coefficients of $\sigma_i^\alpha(t)$, instead of $\rho(t)$, is propagated in exactly the same way as for average purity~\cite{foot0}, that is, because $\Phi=A_I^{-1} \Phi_I$, we have $\Phi_O = A_O A_I^{-1} \Phi_I$. The OTOC vector $\Phi_O$ averaged over single-site random unitaries is therefore propagated as 
\begin{equation}
\Phi_O(t) = M \Phi_O(t-1), \quad M = A_O A_I^{-1} M' A_I A_O^{-1},
\label{eq:OTOC_MC}
\end{equation}
where $M'$ is the transfer matrix propagating purities.

Using $M'$ calculated for random circuits and an arbitrary $W_{i,j}$ parameterized by $(a_{\rm x},a_{\rm y},a_{\rm z})$, Eq.(13) from Ref.~\cite{prejsnji_clanek}, we immediately get the transfer matrix $M$ describing the evolution of average OTOC under one elementary step,
	\begin{equation}
		M_{i,j} =
		\begin{pmatrix}
			1 & 0 & 0 & 0 \\
			0 & c_+ & c_- & d \\
			0 & c_- & c_+ & d \\
			0 & -4d/3 & -4d/3 & (2d+v)/3
		\end{pmatrix}.
		\label{eq:M_cd}
	\end{equation}
Here $c_\pm = \frac{1}{12} \left( 9\pm 2u-v\right)$ and $d=\frac{1}{6} (v-3)$, with $u = \cos\left( \pi \ax \right)+\cos\left( \pi \ay \right)+\cos\left( \pi \az \right)$ and $v = \cos\left( \pi \ax \right)\cos\left( \pi \ay \right)+\cos\left( \pi \ax \right)\cos\left( \pi \az \right)+\cos\left( \pi \ay \right)\cos\left( \pi \az \right)$. Note that $\textbf{e}_0$ is a trivial eigenvector of $M$ corresponding to the eigenvalue $1$, i.e., is a stationary state. However, $M$ has another nontrivial eigenvector with $\lambda=1$ containing the asymptotic stationary values of OTOC $O_\infty$. 

Summarizing, the matrix $M_{i,j}$ (\ref{eq:M_cd}) acts nontrivially only on 2 sites $i$ and $j$ and is written in the basis of bit strings ordered as $\{0_i0_j,1_i0_j,0_i1_j,1_i1_j\}$. To get the matrix propagating OTOCs for the complete circuit for one unit of time one must multiply appropriate 2-site $M_{i,j}$ in the same order as the gates are applied in the protocol, for instance, for a $n=4$ site BW PBC circuit one has $M=M_{4,1} M_{2,3} M_{3,4} M_{1,2}$. The state $\Phi_O$ on which full $M$ acts has $2^n$ components. Fixing the site $i$ in the initial OTOCs $O(i,j,0)$, the initial vector (\ref{eq:OTOC_initial}) is equal to $\Phi_O(0)=(1_0,0_1,\ldots,0_{k-1},\frac{4}{3}_{k},0_{k+1},\ldots,0_{2^n-1})$, where $k=2^{i-1}$ and we number components of $\Phi_O$ starting with $0$. OTOC $O(i,j,t)$ is then equal to $2^{j-1}$-th component of the iterated vector, that is $O(i,j,t)=[M^t \Phi_O(0)]_{2^{j-1}}$.
	
	The transfer matrix description of the average OTOC dynamics (\ref{eq:OTOC_MC}) that we obtained offers several advantages. First, it gives a neat analytical description on which one can use standard tools of analyzing Markov chains, like for instance trying to connect the spectral properties of $M$ to the asymptotic relaxation of OTOC to its infinite-time values. Second, it also greatly simplifies numerical simulations of OTOCs -- instead of, e.g., explicitly simulating the dynamics of operators, averaging over different realizations, one can directly simulate the average OTOCs dynamics.

	\section{Exact dynamics on the light-cone}
\label{sec:LC}		
\begin{figure}[t]
		\begin{center}
			\includegraphics[width=85mm]{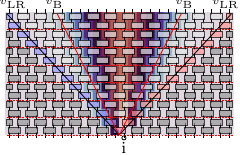}
			\caption{Operator spreading in a random quantum circuit with the brick-wall configuration and the two-qubit gate $W$ with $\ax=0.5$, $\ay=0.3$ and $\az=0.1$ (\ref{eq:parametrizacija}). Background colors represent values of $O(i,j,t)$ -- light gray denotes $O(i,j,t)\approx 0$ and red colors $O(i,j,t)\approx 1$. The operator $\sigma_i^\alpha$ is initially located on the 14th qubit marked by a circle and horizontal dashed red lines are at integer $t$.}
			\label{fig:v_LR}
		\end{center}
	\end{figure}

	 In this section we will obtain the exact dynamics of OTOCs on the light-cone, from which we will be able to determine in a very simple way the set of two-site $W$ that result in maximum velocity circuits. Maximum velocity quantum circuits were defined in Ref.~\cite{maximum_velocity} as circuits where the butterfly velocity $\vB$ \cite{vB_1,vB_2,roberts15,ballistic} equals the Lieb-Robinson velocity $\vLR$ \cite{LiebRobinson}. The Lieb-Robinson velocity determines the causality light-cone of which boundaries are at positions $k = i \pm \vLR t$ ($i$ is the location of $\sigma_i^\alpha(t=0)$), so that OTOC $O(i,j,t)$ with parameters $(j,t)$ outside the light-cone vanish. In a random circuit its value is determined solely by the circuit geometry and is for instance $\vLR=2$ for the BW configuration. The butterfly velocity on the other hand is determined as $\vB=|j-i|/t_{\rm min}$, where $t_{\rm min}$ is the minimal time when $O(i,j,t) \sim 1$ at fixed large $|j-i|$. Contrary to $\vLR$ the butterfly velocity depends both on the geometry and on the choice of the gate $W$. For most random quantum circuits, $\vB \neq \vLR$. An illustration of these two velocities can be found in Fig.~\ref{fig:v_LR}. 

	Let us focus on a random quantum circuit with a brick-wall configuration of gates acting on an infinite system, $n\rightarrow \infty$. We would like to compute quantities $O(i,i \pm \vLR t,t)$. Due to symmetry it is enough to consider OTOC only at the right light-cone. We also limit ourselves to odd $i$ such that the light-cone edge is at $k(t)=i + 2 t$ (for even $i$ one would have $O(i,i+2t,t)=0$ while $O(i,i+2t-1,t)\neq 0$). Remember that in our Markov chain picture $O(i,k(t),t)$ is equal to the $2^{k(t)-1}\equiv(0,\ldots,0,1_k,0,\ldots,0)$-th component of $\Phi_O(t)$. We can get such $O(i,k(t),t)$ by acting with the relevant $M_{k(t)-1,k(t)}$ (red boxes in Fig.~\ref{fig:v_LR}) on a previous half-step $\Phi_O(t-1/2)$. Taking into account that $M_{k(t)-1,k(t)}$ can change only bits at site $k(i)$ and $k(i)-1$, the 3rd row of $M$ (\ref{eq:M_cd}) gets us	
	\begin{align}
		\label{eq:cp}
		O&(i,k(t),t) = c_- [\Phi_O(t-1/2)]_{2^{k(t)-1}}\, +   \\	
		& + c_+ [\Phi_O(t-1/2)]_{2^{k(t)}} + d [\Phi_O(t-1/2)]_{2^{k(t)-1}+2^{k(t)}}.\nonumber
	\end{align}
	It is important to note that due to causality all values $[\Phi_O(t)]_{p}$ with $p \ge 2^{k(t)}$ vanish, therefore only one term in Eq.~(\ref{eq:cp}) is nonzero, resulting in $O(i,k(t),t) = c_- O(i,k(t)-1,t-1/2)$. Iterating this by half-steps to smaller times until we reach $O(i,i,0) = 4/3$ one obtains the OTOC on the right light-cone. Similar procedure works also on the left light-cone and even $i$, resulting in
		\begin{equation}
			O(i,i\pm 2 t,t) = \frac{4}{3} (c_-)^{2t}.	
			\label{eq:OTOC_lcr}
		\end{equation}
	Looking at the left light-cone at odd $i$, or the right light-cone and even $i$, one instead gets
		\begin{equation}
				O(i,i\pm(2 t-1),t)= \frac{4}{3} c_+(c_-)^{2t-1}.
				\label{eq:OTOC_lcl}
		\end{equation}
	The additional term $c_+$ in Eq.~(\ref{eq:OTOC_lcl}) comes from the interaction at time $t=1/2$, namely $O(i,i,1/2) = c_+ O(i,i,0)$.
	
	OTOC on the light-cone therefore decay exponentially with the rate $2\ln{c_-}$, hence one gets $\vB=\vLR=2$ iff $c_-=1$. Solving $c_-=1$ for $\ax,\ay,\az$ we obtain $\ax=\ay=1$ and an arbitrary $\az$, which corresponds to dual-unitary circuits~\cite{DU-LC} and for which one can explicitly calculate all 2-point correlations which are nonzero only on the light-cone boundary~\cite{DU-LC}. This means that taking $W$ from the dual-unitary set of gates (i.e., so-called XXZ gates) is the only choice leading to the maximum velocity random circuits (of the type studied in this paper), i.e., circuits for which OTOC do not decay along the light-cone. The same set of maximum velocity gates was also obtained in Ref.~\cite{maximum_velocity} for circuits without one-site random unitaries. Besides identifying maximum velocity gates our simple derivation also gets us the exact dynamics of OTOC on the light-cone for arbitrary gates $W$. Note that for circuits with a dual-unitary two-qubit gate $W$ one can also get a closed expression for the OTOC decay in the vicinity of the light-cone, for non-random circuits see~\cite{maximum_velocity}, for random~\cite{Bruno20}. 

We also observe that the same set of gates, except at $\az=1$, results in the maximal possible entanglement scrambling speed~\cite{prejsnji_clanek}. The gate with $\az=1$ is the SWAP gate and is special. The OTOC dynamics for a random circuit with the SWAP gate is trivial because the transfer matrix $M_{i,j}$ itself (\ref{eq:M_cd}) is equal to a SWAP gate resulting in $O(i,j,t)$ that is non-zero only on the light-cone, while at the same time such $W$ produces no entanglement.

	\section{Convergence rate}
\begin{figure}[t]
		\begin{center}
			\includegraphics[width=85mm]{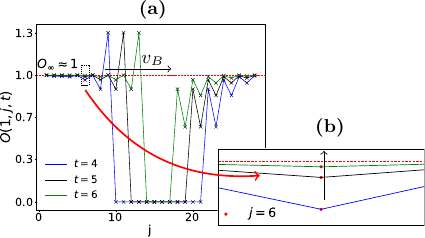}
			\caption{Average OTOC for the BW random circuit with PBC and the XY gate ($\textbf{a} = (1,1,0)$), $n = 28$. Operator relaxation towards long-time thermal asymptotics will be studied by observing how a fixed-position $O(i=1, j = 6, t)$ converges towards $O_{\infty}$ with time (inset).}
			\label{fig:OTOC_convergence}
		\end{center}
	\end{figure}

	Under the application of a random quantum circuit the initially localized operator will spread in space, causing OTOC to increase from being zero outside of a light-cone to a nonzero value inside it. Often one is interested in this ramp-up of OTOC as for instance measured by the butterfly velocity. We shall instead investigate the late-time convergence rate of OTOC $O(i,j,t)$. That is, we are interested in how fast $O(i,j,t)$ at some fixed $i$ and $j$ relaxes towards its final value, see Fig.~(\ref{fig:OTOC_convergence}) for an illustration.

The asymptotic value $O_\infty$ is reached at long times when the time evolved operator $\sigma_i^\alpha(t)$ becomes a uniform mixture of all possible Pauli strings (identity excluded) on $n$ qubits, i.e., when the propagator $U$ resembles a random unitary. A derivation of $O_\infty$ can be found~\cite{yoshida17} in \ref{app:O_inf} and gives us 
   \begin{equation}
	   O_\infty  = 1+\frac{1}{4^n-1}.
	   \label{eq:OTOC_inf}
   \end{equation}
If the eigenvalues of the transfer matrix $M$ are gapped away from $1$, which indeed is the case, we expect that OTOC exponentially relax to their asymptotic value $O_\infty$ as 
	\begin{equation}
		|O(i,j,t)-O_\infty|  \asymp \mathrm{e}^{-r t}. \label{eq:OTOC_asymp}
	\end{equation}
Our main object of study is the convergence rate $r$. Considering that OTOC are propagated by $M^t$ one might think that the convergence rate will be determined by the 2nd largest eigenvalue $\lambda_2$ of $M$. However, as we shall see, this is not always the case.

In section \ref{sec:random} we shall first discuss protocols in which at each step we randomly pick a pair of qubits on which we act, that is protocols with a random ordering of gates. For those we will see that indeed OTOC decay exponentially with the convergence rate $r$ given by the second largest eigenvalue of the transfer matrix $r = -\ln|\lambda_2|$.

In sections \ref{sec:pbc} and \ref{sec:obc} we shall on the other hand study protocols with a nearest-neighbor deterministic order of gates, mostly the BW or the S configuration (Fig.~\ref{fig:BWandS}), and find, similarly as for purity~\cite{prejsnji_clanek}, that the convergence rate can be either equal or smaller than $-\ln|\lambda_2|$. In Appendix~\ref{app:randomness} we also numerically demonstrate that in the thermodynamic limit one does not need explicit averaging over single-qubit unitaries, i.e., dynamics is self-averaging and therefore one will obtain the same results also for a single circuit realization.

Relying on a map from OTOC to a partition function of an Ising-like model found in \cite{adam18} we analytically compute OTOC for the BW PBC and BW OBC in the case where every elementary step of the circuit is independently drawn from the Haar measure on U(4). Contrary to previous literature, where the analytic expression for OTOC was obtained in the TDL \cite{adam18,maximum_velocity,Bruno20}, in Appendix~\ref{app:U4_results} we present new results in finite systems with either OBC or PBC.
	
	\subsection{Random protocols}
\label{sec:random}	

	Random protocols studied here are defined as random quantum circuit where at every elementary step we couple two qubits chosen randomly. We have two different possibilities: a) at each step we uniformly choose one of the all possible $n$ qubits, for example the $i$-th one, and we act with gates $M_{i,i+1}$, and b) at each step we randomly choose two qubits $i$ and $j$ and we act with $M_{i,j}$. We call the former case the random nearest neighbor protocol (r.n.n.) and the latter scenario the all-to-all coupling.
	
	In the r.n.n. case, the average elementary step can be written as the average over all possible choices of $i$, namely
	\begin{equation}
		\bar{M} = \frac{1}{L} \sum_{i=1}^L M_{i,i+1},
	\end{equation}
with $L=n-1$ or $L=n$, depending on the boundary conditions. Similarly, for the all-to-all case we obtain
	\begin{equation}
		\bar{M} = \frac{1}{L} \sum_{i<j} M_{i,j},
	\end{equation}
with $L=n(n-1)/2$. The transfer matrix propagating OTOC for one unit of time is $M=\bar{M}^{L}$.

Because each $M_{i,j}$ is just a similarity transform of the purity $M'_{i,j}$, Eq.~(\ref{eq:OTOC_MC}), the spectrum of $M$ is identical to the spectrum of purities transfer matrix $M'$. Furthermore, as was shown in \cite{prejsnji_clanek}, the average elementary steps $M'_{i,j}$ propagating purity can be linearly transformed to a real symmetric matrix. Therefore the spectrum of $M$ is equal to the spectrum of a symmetric purity matrix. This is important because the spectrum is real with orthogonal eigenvectors. The spectral decomposition of a Hermitian $M$ takes the form $M = \sum_k \lambda_k \ket{v_k} \bra{v_k}$, and we can expand $\ket{\Phi}$ as $\ket{\Phi} = \sum_k c_k \ket{v_k}$ with $c_k = \braket{v_k}{\Phi}$ being bounded by $|c_k|^2 \le \braket{\Phi}{\Phi}$. The time iteration is $\Phi(t)=M^t \Phi$ from which it follows that $|\Phi(t)-\Phi(t\rightarrow \infty)| \asymp |\lambda_2|^t$ with $\Phi(t\rightarrow \infty) = v_1$. 

For a Hermitian $M$ there are therefore no surprises; if the 2nd largest eigenvalue $\lambda_2$ is gapped away from other eigenvalues the asymptotic decay rate will be given by $\lambda_2$ and will kick-in at a system size independent time. For the r.n.n. protocol the 2nd largest eigenvalue has been computed numerically for arbitrary gates \cite{Znidaric_2007} and analytically for a few Clifford gates \cite{PRA08}. For $\lambda_2$ in the all-to-all case and Clifford $W$ see Ref.~\cite{PRA08}, for arbitrary $W$ Ref.~\cite{prejsnji_clanek}.

	\subsection{Brick-wall protocol with PBC}
\label{sec:pbc}

For protocols with a deterministic order of gates things can and will be completely different. The decay rate will not necessarily be given by $\lambda_2$. Remember that for a deterministic order of gates the transfer matrix is just a product of corresponding two-site $M_{i,j}$, for instance, for a 4 qubit BW protocol it is $M=M_{4,1} M_{2,3} M_{3,4} M_{1,2}$. The difference compared to random protocols is that a product of symmetric matrices needs not to be symmetric. As a consequence, the eigenvectors of such $M$ are not orthogonal, $c_k$ are not upper bounded, and, as has been seen in purity evolution \cite{prejsnji_clanek}, the relevant decay can differ from $|\lambda_2|^t$.

    In the following we shall plot how the value $O(1,j,t)$ behaves for a fixed position $j$. We will always fix $i=1$, because OTOC in PBC circuits depend only on $|j-i|$. We will plot values of $|O(1,j,t)-O_\infty|$ and the time derivative 
	\begin{equation}
		r(t) := -\frac{\mathrm{d}}{\mathrm{d}t} \ln|O(1,j,t)-O_\infty|
	\end{equation}
	in order to investigate OTOC convergence rate (note that $r(t\rightarrow \infty) = r$ from Eq.~\ref{eq:OTOC_asymp}).
	
\begin{figure}[t]
		\begin{center}
			\includegraphics[width=85mm]{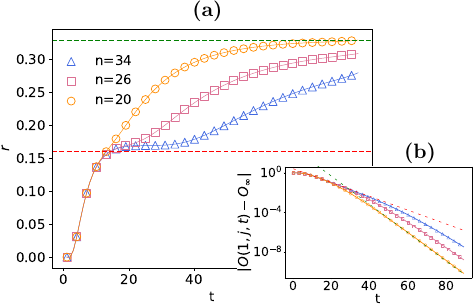}
			\caption{Convergence rate of $O(1,j=7,t)$ for a BW PBC circuit with the gate $W_g$ (Eq.~\ref{eq:Wg}). There is a phantom eigenvalue: initially, the rate is given by $\lambda_2$ for a BW OBC circuit (red dashed line for $n=30$). At late times the rate is instead equal to $-\ln|\lambda_2|$ for a BW PBC circuit (green dashed line for $n=20$). The inline plot shows a transition in the exponential decay of the same data, including red and green dashed exponential functions corresponding to red and green rates in the main plot.} 
			\label{fig:PBC_BW_nondual}
		\end{center}
	\end{figure}

	Let us start with a generic two-qubit gate 
	\begin{equation}
		W_g = W(\textbf{a}), \quad \textbf{a}=(0.5,0.3,0.1).
		\label{eq:Wg}
	\end{equation}
Data for $O(1,j=7,t)$ is shown in Fig.~\ref{fig:PBC_BW_nondual} and demonstrates that OTOC converge to their final value with a rate different than $-\ln|\lambda_2|$ (for $W_g$ one has $|\lambda_2|\approx0.72$ for $n=20$). The rate is (initially) smaller, as if there would be an eigenvalue larger than $\lambda_2$ - a phantom eigenvalue. Such slower decay persists up to times that are proportional to the system size. The value of the phantom eigenvalue is equal to the second largest eigenvalue of the transfer matrix for the BW OBC circuit. Remember that we are looking at a circuit with PBC, not OBC, nevertheless, it is perhaps expected that for initially localized quantities and until the boundary conditions (PBC) influence OTOC dynamics, the convergence rate is given by $\lambda_2$ of BW OBC (see also next section). Namely, choosing the initial vector localized roughly equally far from the left and right boundary ($i\approx n/2$), the dynamics generated by BW PBC or OBC circuit is identical up to times $t\approx n/4$. Therefore, what might be surprising is that $\lambda_2$ of $M$ for BW with PBC and OBC are different. Looking at OTOC on a different site, $j\neq 7$, one might observe a slightly different graph from Fig.~\ref{fig:PBC_BW_nondual}, however the behavior remains qualitatively the same: at early times that scale as $\sim n$ the dynamics is always determined by a phantom eigenvalue, which is the same for every $j$, whereas at late time the dynamics is given by the second largest eigenvalue of $M$.

Of special interest are gates with canonical parameters $\textbf{a} = (1,1,\az)$, $\az < 1$ (dual unitary $W$). We purposely skip $\az=1$ because of its trivial dynamics. Contrary to the generic gates $W_g$, for dual unitary gates we will see that early-time dynamics is always determined by $\lambda_2$ of an S PBC circuit (shown in Fig.~\ref{fig:BWandS}(b)), which though is always larger than $\lambda_2$ for BW PBC. One will therefore again have a situation where the relevant relaxation rate is not given by $\lambda_2$ of the BW PBC circuit.

\begin{figure}[t]
		\begin{center}
			\includegraphics[width=80mm]{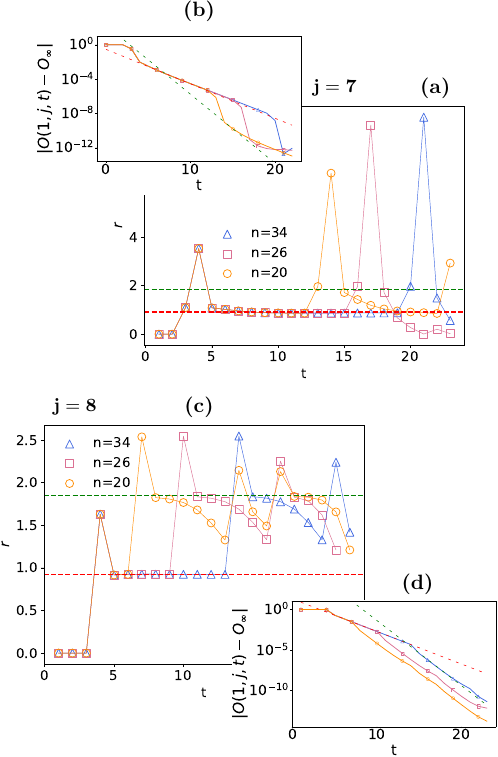}
			\caption{Convergence rate $r$ for the BW PBC circuit with $\textbf{a}=(1,1,0.2)$. Dynamics up to times $t\sim n$ is determined by the second largest eigenvalue of the transfer matrix for the S PBC configuration (red dashed line), whereas the late time dynamic is determined by $|\lambda_2|$ of the BW PBC (green dashed line).}
			\label{fig:PBC_BW_az02}
		\end{center}
	\end{figure}

    We will first take a look at a circuit with the dual-unitary gate with $\az=0.2$. Because there are some differences between even and odd $j$ at later times, essentially due to even/odd effects of the light-cone boundary position (see Sec.\ref{sec:LC}), we show in Fig.~\ref{fig:PBC_BW_az02} how $O(1,j,t)$ converge for $j=7$ in (a) and (b), as well as for $j=8$ in (c) and (d). Looking at Fig.~\ref{fig:PBC_BW_az02}(c) that focuses on short times we can see that $r$ is zero until the right light-cone boundary hits the site $j=8$. We assume that $j-i$ is odd and $j-i<n-(j-i)$, i.e., the first information that hits the site $j$ comes from the right light-cone boundary, not from the wrapped-around (PBC) left light-cone boundary. OTOC and the rate are therefore zero until $t \approx (j-i)/2$. After that $r$ stays at a value that is not given by $|\lambda_2|$ of the BW PBC transfer matrix, but rather by $\lambda_2$ of the transfer matrix for the S PBC (red dashed line) and for which we have a conjectured analytical form, see Ref.\cite{foot1}). At the time $t_c = (n+1)/2-(j-i)/2$, determined by the time when the left light-cone boundary hits the site $j$, the rate suddenly transitions to its ultimate asymptotic value given by $\lambda_2$ of the BW PBC (green line). In Fig.~\ref{fig:PBC_BW_az02}(d) we can see that this rate stays roughly constant upto small modulations at times larger than $t_c$, e.g. at $t\approx 20$ for $n=34$. They happen at times of successive light-cone boundary wrappings (for more details see next paragraph). There are some interesting differences for odd $j$ (Fig.\ref{fig:PBC_BW_az02}(a,b)). Specifically, because for odd $i=1$ the left light-cone boundary is at even sites and therefore never overlaps with an odd $j$, the rate has a transition to its asymptotic form only when the right light-cone boundary hits the odd site $j=7$ for the 2nd time (due to PBC). This happens at $t_c =(j-i)/2+n/2$, e.g., $t_c=20$ for shown $n=34$ and $j=7$. As one can see from $O$ in Fig.\ref{fig:PBC_BW_az02}(b), the rate itself does not change; rather the OTOC exhibits a jump. As we shall see in the next paragraph, the ultimate asymptotic decay is nevertheless still determined by $\lambda_2$ of the BW PBC circuit.

\begin{figure}[t]
 		\begin{center}
 			\includegraphics[width=80mm]{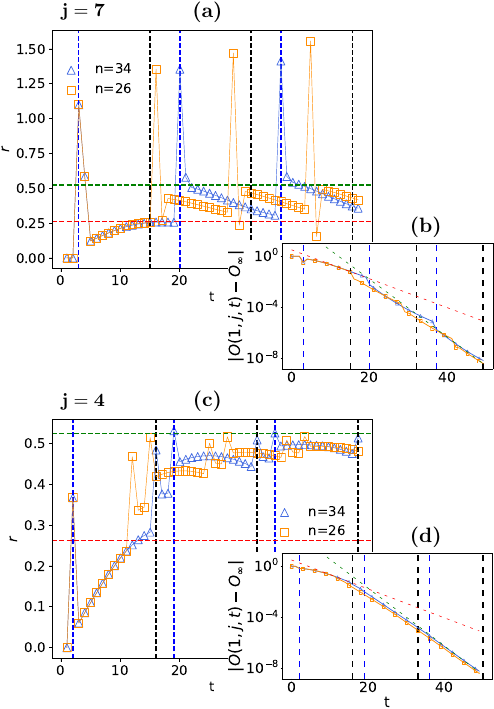}
 			\caption{Time evolution of OTOC convergence rate for BW PBC circuits with $\textbf{a}=(1,1,0.6)$. Red and green dotted lines represent the convergence rate determined by $\lambda_2$ of S PBC and BW PBC circuits respectively. When $j$ is odd ((a) and (b)) spikes in $r$ are found at times when the right light-cone boundary hits site $j$, $i+2t=j \pmod{n}$ (blue vertical dashed lines plotted for $n=34$). For even $j$ ((c) and (d)) these spikes can be found also at $i-2t+1=j \pmod{n}$ when the left light-cone boundary hits site $j$ (black vertical dashed lines for $n=34$).}
 			\label{fig:PBC_BW_az06}
 		\end{center}
 	\end{figure}

In order to be able to better explore those spikes we shall next look at a dual unitary gate with $\az=0.6$, because OTOC decay slower and we are able to simulate longer times (rounding errors of double precision floating point numbers ultimately limit the smallest $O$ we can calculate). Results are shown in Fig.~\ref{fig:PBC_BW_az06}. From the figure we learn that the convergence of the initial rate to that given by the eigenvalue of the S PBC protocol is rather slow with $n$; smaller system sizes have rates that do not yet converge to $-\ln|\lambda_2|_{\mathrm{S-PBC}}$. There are also small kinks in the decay of $|O(1,j,t)-O_\infty|$ that are due to light-cone wrapping boundaries. Overall though the rate changes only once from the initial one to the asymptotic $-\ln|\lambda_2|_{\mathrm{BW-PBC}}$ at the already discussed time that is proportional to $n$ (see frames (b) and (d)). We can now also clearly see several spikes at times when the light-cone boundary wraps around the system multiple times. Specifically, starting with an odd $i=1$ the right light-cone boundary will hit a site at odd $j$ at times $t=(j-i)/2+k n/2$, where $k$ in an integer (blue vertical lines in the Figure), whereas the left light-cone boundary will hit it at times $t=k n/2-(j-i-1)/2$ (black vertical lines). There is a slight asymmetry between the effects of left and right light-cone boundary: spikes due to the left one are prominent only for even $j$ which comes due to an asymmetry in the behavior of OTOC on the light-cone boundary, Eqs.(\ref{eq:OTOC_lcr}\ref{eq:OTOC_lcl}) -- for even $j$ the left light-cone has an additional factor $c_+=\frac{1}{3}(1+\cos{(\pi \az)})=1-|\lambda_2|_{\rm S-PBC}$.

We also observe that $r$ decreases with increasing $\az$ (\cite{foot1}). This means that the fastest relaxation of OTOC among dual-unitary gates is obtained for the circuit with the XY gate, i.e., $\textbf{a}=(1,1,0)$, when one has $r=\log{3}$ at $t \lesssim n$ .
	
\bigskip
	
We have seen that in all BW circuits with PBC, for generic gates as well as dual unitary gates, the relevant relaxation rate of OTOC that holds upto times of order $\sim n$, i.e., until OTOC become exponentially small in system size, is not given by the 2nd largest eigenvalue of the BW PBC transfer matrix. For the generic gate $W_g$ the rate was given by $|\lambda_2|_{\mathrm{BW-OBC}}$ which is larger than $|\lambda_2|_{\mathrm{BW-PBC}}$ -- a phantom eigenvalue phenomenon. One might be inclined to justify this result based on a trivial fact that the choice of boundary conditions of course does not matter up to times that are proportional to $\sim n$. Until boundary effects kick in OTOC evolve as they would in a BW OBC system (if the Pauli matrix at time $t=0$ is positioned ``far enough" from the boundaries). This however is not really a full explanation; remember also that for dual unitary gates the rate (phantom eigenvalue) was given by $|\lambda_2|_{\mathrm{S-PBC}}$ and not $|\lambda_2|_{\mathrm{BW-OBC}}$, despite the evolution still being the same as it would be in the BW OBC circuit.

In the next section we shall study circuits with OBC. Based on results presented so far we can predict that for BW OBC with generic gates one will have no phantoms, whereas we expect to see a phantom rate given by $|\lambda_2|_{\mathrm{PBC-S}}$ for BW OBC circuits with dual unitary gates.

	\subsection{OBC protocols}
\label{sec:obc}	
Let us first stress one important property of a family of OBC protocols that comes about due to the locality of the initial vector $\Phi_O(t=0)$ (Eq.~\ref{eq:OTOC_initial}). We conjecture that OTOC dynamics is not influenced by permutations of elementary gates in one period of the BW OBC circuit. For example, looking at a BW OBC protocol one could permute the order of elementary steps in one period and obtain an S OBC circuit without affecting the average OTOC dynamics, e.g., its decay rate.
	\begin{figure}[t]
		\begin{center}
			\includegraphics[width=50mm]{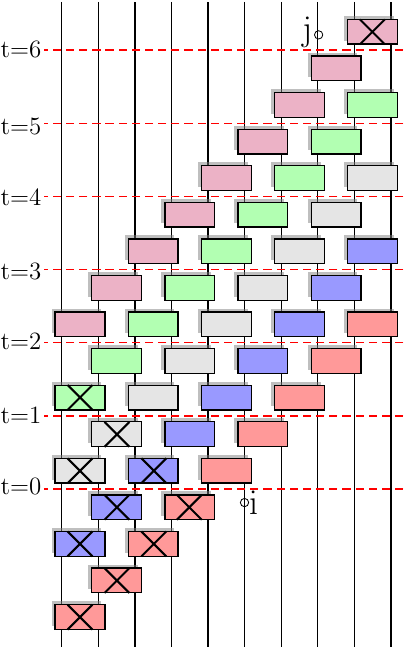}
			\caption{Comparison between $O(6,8,t)$ obtained using 5 iterations of a S OBC circuit and using a BW OBC circuit. Operators in the same period of the S OBC circuit are represented with the same color, meanwhile operators in the same period of BW OBC are labeled by the same parameter $t$ (see Fig.~\ref{fig:BWandS}). Due to causality, all gates outside the future light-cone starting from $i=6$, and the past light-cone originating from $j=8$, do not matter (crossed out gates). By stacking together S OBC protocols one obtains the same set of gates as for BW OBC.}
			\label{fig:BW_is_S}
		\end{center}
	\end{figure}

\begin{figure}[t]
		\begin{center}
			\includegraphics[width=80mm]{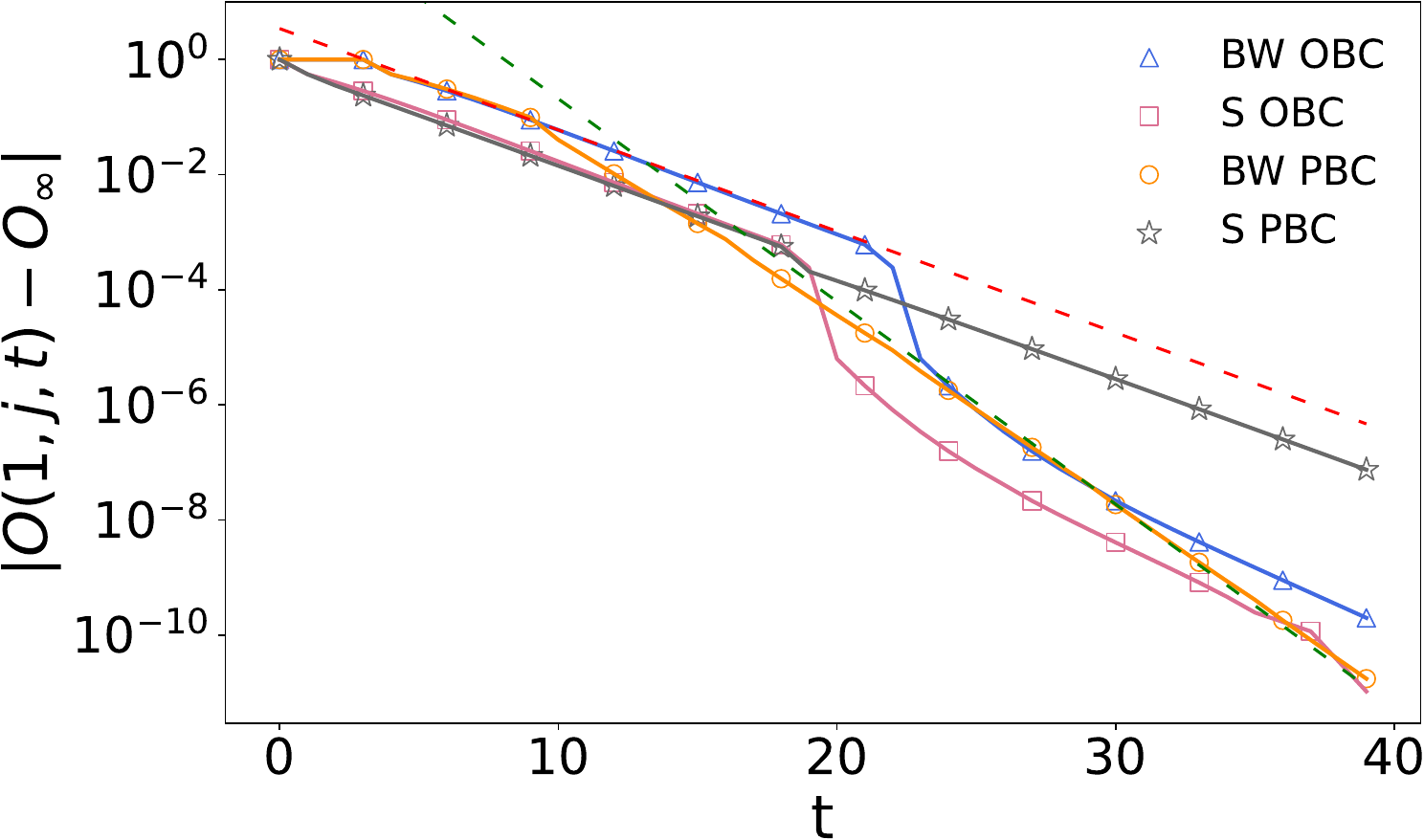}
			\caption{Comparison of OTOC relaxation for different protocols and $\textbf{a}=(1,1,0.5)$, $i=1$, $j=8$ and $n=26$. BW OBC and S OBC are equivalent up to a time-shift (equal to $3$ in this case). For PBC on the other hand S and BW, while having the same initial decay, exhibit different relaxation rate at long times. The asymptotic decay of BW PBC is given by $|\lambda_2|$ of the BW PBC (green dashed line), while that of S PBC it is given by $|\lambda_2|$ of the S PBC (red dashed line).}
			\label{fig:diff}
		\end{center}
	\end{figure}

        To support this claim, we will show that for a fixed $W$ the BW OBC protocol generates the same OTOC dynamics as the S OBC protocol upto a constant time-shift. We  will rely on Fig.~\ref{fig:BW_is_S} to explain the equivalence. One can easily see that stacking together $S$ protocols one obtains a circuit of the form shown in Fig.~\ref{fig:BW_is_S}, i.e., a brick wall protocol in the middle (between $t=2$ and $t=3$ in the Figure), and two ``triangles", one at the top right (gates after time $t=3$ in Fig.~\ref{fig:BW_is_S}) and one at bottom left (gates before time $t=2$ in Fig.~\ref{fig:BW_is_S}). Let us focus on the calculation of $O(i,j,t)$. Due to causality the evolved local operator vanishes outside the light-cone starting from the $i$-th qubit. We are interested in the component of $\Phi_O(t)$ representing $O(i,j,t)$, i.e. the $2^{j-1}$-th one, this means that also operators in the past light-cone starting from the qubit $j$ vanish. The relevant gates are therefore those inside the two light-cones, i.e., in Fig.~\ref{fig:BW_is_S} the gates that are not crossed. The same set of relevant gates would be obtained acting with a BW OBC protocol. The only difference between S OBC and BW OBC circuits is a time-shift that comes from the difference between $\vLR$ of the two circuits. This is reflected in the fact that $O(i,j,1) \neq 0$ for arbitrary $j$ in S OBC circuits, whereas using a BW OBC protocol we have $O(i,j,t) = 0$ at all times smaller than $\Delta t \approx |j-i|/2$, which is equal to the time-shift between the two protocols. For instance, by counting the number of BW layers of relevant gates in Fig~\ref{fig:BW_is_S} one can see that $O_{\rm S}(6,8,5)=O_{\rm BW}(6,8,6)$. The time-shift is constant and depends only on the value $j-i$ (and can be a half-integer). This can be seen also in explicit numerical data in Fig.~\ref{fig:diff} where $O(1,8,t)$ for BW OBC circuit (triangles) is the same as $O(1,8,t-3)$ (squares) obtained for the S OBC. 

Using similar arguments one can see that if one iterates an arbitrary OBC configuration, that is a protocol in which each nearest-neighbor gate is applied exactly once per unit of time, one always gets a brickwall pattern of gates. Therefore one can show that the OTOC of local operators and any OBC protocol is upto a time-shift equal to the one in say BW OBC circuit. We have also checked numerically on a few examples of random gate permutations that this is indeed the case. We remark that in Ref.~\cite{prejsnji_clanek} it has been shown that the spectra of transfer matrices $M$ for a single iteration are the same for all OBC protocols.

This equivalence though holds only for OBC. For instance, for the XXZ gate S PBC and BW PBC protocols can behave rather differently, see circles and stars in Fig.~\ref{fig:diff}, what is more, the BW PBC circuit exhibits a phantom relaxation. On the other hand, the S PBC with the XXZ gate does not exhibit a phantom relaxation, while the S PBC with the generic gate $W_{\rm g}$ does (data not shown).
	
\begin{figure}[t]
		\begin{center}
			\includegraphics[width=80mm]{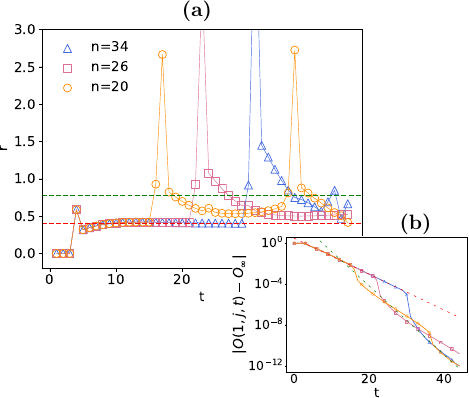}
			\caption{Convergence rate for BW OBC circuits with $\az=0.5$, $j=8$. Red and green dashed lines denote the rate predicted by $|\lambda_2|$ for S PBC and BW OBC circuits respectively. Spikes in the rate are again found at times when the light-cone boundary is reflected from boundaries back to site $j$.}
			\label{fig:OBC_BW_az05}
		\end{center}
	\end{figure}

Regarding possible phantoms in the OBC setting we can see in Fig.~\ref{fig:diff} that for dual unitary gates BW OBC does exhibit a phantom (the initial rate is given by $|\lambda_2|$ of the S PBC), while for generic gates it does not (data not shown), which is expected (we have seen in Fig.~\ref{fig:PBC_BW_nondual} that the rate for BW PBC was given by $|\lambda_2|$ of BW OBC, and the two $O(i,j,t)$ should agree until $t \sim n$). Let us have a closer look at the dual unitary gate $\az=0.5$ and BW OBC protocol. From data in Fig.~\ref{fig:OBC_BW_az05} we indeed see that there is a phantom -- the initial rate is smaller -- and that there are, similar as in the PBC case (Fig.~\ref{fig:PBC_BW_az06}), again spikes in the rate. Those spikes are associated with jumps in the relaxation of OTOC (frame (b)) that happen every time the reflected right light-cone returns to site $j$, i.e., at times $kn-(j-i-1)/2$. 

\begingroup
\squeezetable
\begin{table}[t!]
\begin{ruledtabular}
\begin{tabular}{ccccccc}
\multicolumn{1}{c}{Gate}  & \multicolumn{2}{c}{Protocol} & \phantom{a} & Phantom   & True eig. \\

\cmidrule(r){2-3}
& config. & b.c. & & $\lambda_{\rm ph}$ & $|\lambda_2|$\\
\midrule

                  & S   & OBC && no              & $0.86=|\lambda_2|_{\rm obc}$ &\\
$W_g$             & S   & PBC && $0.86=|\lambda_2|_{\rm obc}$  & $0.74$ &\\
$\mathbf{a}=(0.5,0.3,0.1)$   & BW  & OBC && no              & $0.86=|\lambda_2|_{\rm obc}$ &\\
                  & BW  & PBC && $0.86=|\lambda_2|_{\rm obc}$  & $0.73$ &\\
\multicolumn{1}{l}{}\\
                  & S   & OBC && $\frac{2}{3}=|\lambda_2|_{\rm Spbc}$ & $0.45$ &\\
 XXZ              & S   & PBC && no & $\frac{2}{3}=|\lambda_2|_{\rm Spbc}$ &\\
e.g. $\mathbf{a}=(1,1,0.5)$  & BW  & OBC && $\frac{2}{3}=|\lambda_2|_{\rm Spbc}$   & $0.45$ &\\
                  & BW  & PBC && $\frac{2}{3}=|\lambda_2|_{\rm Spbc}$ & $\frac{4}{9}=|\lambda_2|_{\rm BWpbc}$          &\\

\end{tabular}
\end{ruledtabular}
\caption{Phantom eigenvalue $\lambda_{\rm ph}$ and the second largest eigenvalue $|\lambda_2|$ for different random circuits. For the generic gate $W_{\rm g}$ one has phantom relaxation in both PBC cases, while for the XXZ gate in addition also for the S OBC protocol (remember that for the OBC the spectrum does not depend on the configuration~\cite{prejsnji_clanek}, i.e., $\lambda_2$ is the same for BW and S). The 2nd largest eigenvalue for XXZ gates and PBC protocols are conjectured to be equal to $|\lambda_2|_{\rm Spbc}=(2-\cos(\pi \az))/3$ and $|\lambda_2|_{\rm BWpbc}=(2-\cos(\pi \az))^2/9$~\cite{foot1}.}
\label{tab:overview}
\end{table}
\endgroup
All different cases of random circuits and their relaxation are summarized in Table~\ref{tab:overview}. Based on numerical values of $\lambda_{\rm ph}$ (for a number of different gates; see also previous Figures) we identify that $\lambda_{\rm ph}$ is equal to either $|\lambda_2|$ for the OBC in the case of generic gates, or to $|\lambda_2|_{\rm S-PBC}$ in the case of XXZ gates. Predicting when does one have a phantom relaxation and what is this $\lambda_{\rm ph}$ equal to is not simple. In some cases $\lambda_{\rm ph}$ is equal to the second largest eigenvalue for a circuit with OBC, like for the generic gate $W_{\rm g}$ with PBC, while for the XXZ gate with S OBC protocol it is instead equal to $|\lambda_2|$ for the PBC. There is also the case of XXZ gates in the BW configuration where the phantom eigenvalue is equal to $|\lambda_2|$ for a whole different circuit, namely the S configuration. Understanding in detail the physics of phantom relaxation therefore remains an open problem.

	\section{Conclusion}
	We have derived a Markovian propagator for the average out-of-time-ordered correlations of local operators in random quantum circuits in which each two-qubit transformation is composed of a fixed two-qubit gate $W$ and two random single-qubit unitaries. This allows us to get an exact expression for OTOC on the light-cone and any $W$. 

We then focus on the asymptotic relaxation rate at long times with which OTOC relaxes to its long-time average corresponding to a completely scrambled evolution. Similarly as in the case of purity relaxation~\cite{prejsnji_clanek} we find that this OTOC relaxation rate is in many cases not given by the second largest eigenvalue of the Markovian matrix $M$ that governs dynamics of OTOCs. One has a so-called phantom relaxation -- a relaxation where the approach to the steady state asymptotically goes as $\lambda_{\rm ph}^t$, with $\lambda_{\rm ph}$ being some number that is, opposite to ``expectations'', not equal to any of the eigenvalues of $M$. Because $\lambda_{\rm ph}$ is in fact larger than any nontrivial eigenvalue $|\lambda_j|$ of $M$ we call it a phantom eigenvalue. In short, $M$ encodes all the information about OTOCs evolution but its eigenvalues do not give the correct relaxation rate in the thermodynamic limit, despite $|\lambda_2|$ being gapped away from $\lambda_1=1$.
 
Such phantom relaxation proceeds in two steps, where in the first step that lasts upto times that are linear in system size the rate is given by the phantom eigenvalue, while in the second it is eventually given by the second largest eigenvalue $\lambda_2$. Because the transition time between the two regimes diverges in the thermodynamic limit one has a situation where at a fixed system size and $t \to \infty$ one gets the naively expected (but thermodynamically incorrect) relaxation as $|\lambda_2|^t$, while in the correct thermodynamic limit of first taking the system size to infinity and only then time to infinity one will observe the relaxation rate given by the phantom eigenvalue. The phenomenon occurs because the limits $t \to \infty$ and $n \to \infty$ do no commute, while mathematically it comes about because the transfer matrix $M$ is not symmetric, resulting in spectral expansion coefficients that blow up with system size~\cite{prejsnji_clanek}, see also Refs.~\cite{Mori20,sarang21,ueda21,mori21,lamacraft21} for other situations where that occurs.

We find such two-step phantom relaxation for brick-wall circuits with dual unitary (i.e., XXZ type) as well as with generic two-qubit gates, and for periodic or open boundary conditions. Phantoms are also found for the staircases configuration with open boundary conditions and dual unitary type gates; see Table~\ref{tab:overview} for an overview. We numerically observe that the phantom eigenvalue $\lambda_{\rm ph}$ is equal to the 2nd largest eigenvalue of $M$ of a different circuit that can have different boundary conditions as well as different gates configuration (theoretical reasons for that are at present not understood). 

For circuits with open boundary conditions we demonstrate that up to a time-shift all different circuit geometries, i.e., brick-wall, staircases, etc., have the same OTOC dynamics. We also numerically verify that the dynamics is self-averaging, that is, one will get a phantom relaxation even for a single random circuit realization, and even without spatial or time independence of single-qubit random unitaries. An explicit randomness therefore seems not to be essential. This leaves an interesting possibility that a similar phenomenon could be observed also in in other systems, for instance in Floquet models.

        The important message therefore is that: (i) when one deals with finite non-Hermitian matrices the leading eigenvalue might not give the correct asymptotic dynamics, and (ii) that this leads to a two-step relaxation process with a sudden discontinuous transition in the relaxation rate at a time when the light-cone hits the site in question for the second time (either due to a reflection from a boundary for open boundaries, or due to a wrapping around for periodic boundary conditions). On the mathematical level it is therefore due to the fact that boundary conditions apparently can affect the leading relevant eigenvalue in a nontrivial way. While we do obtain some exact properties of the Markovian matrix, like a conjectured exact expression for $\lambda_2$ in the case of periodic boundary conditions, much remains to be understood, in particular under which physical conditions one gets such a two-step relaxation.

Support from Grants No.~J1-1698 and No.~P1-0402 from the Slovenian Research Agency is acknowledged.

	\newpage
	
	\appendix
	
	\section{OTOC properties}
	\label{app:OTOC_properties}
	
\subsection{Asymptotic value}
 \label{app:O_inf}
 Here we calculate the asymptotic value $O_{\infty}$ of average OTOC (see Eq.~\ref{eq:OTOC_first_def}) evolved with random quantum circuits, see also Ref.\cite{yoshida17}. After long time, the propagator for the random quantum circuit will resemble a random unitary operator, that is unitary uniformly drawn from the group $\mathrm{U(2^n)}$. Therefore, to get the long-time value of the OTOC we can replace an explicit averaging over circuits, or over long time, with a Haar average over $\mathrm{U(2^n)}$,
 \begin{equation}
  O_{\infty} = 1-\frac{1}{2^n} \mathop{\mathbb{E}}_{\mathrm{U \in 
\mathrm{Haar}}} \tr [ \sigma_i^\alpha(t) \sigma_j^\beta \sigma_i^\alpha(t) \sigma_j^\beta ],
 \label{eq:app_Oinf}
 \end{equation}
where $\sigma_i^\alpha(t)=U^\dagger \sigma_i^\alpha U$. To simplify calculations, we exploit the fact that the Clifford group (unitaries that map Pauli strings into Pauli strings) is a 2-design, so the average over $\mathrm{U(2^n)}$ in Eq.~\ref{eq:app_Oinf} can be replaced by the average over the Clifford group. Note that the $t$ dependence in $\sigma_i^\alpha(t)$ is now superficial as $U$ in our averaging runs over all elements of the Clifford group. For the Clifford $U$ the transformed $\sigma_i^\alpha(t)$ is also a product of Pauli matrices and therefore it either commutes or anticommutes with $\sigma_j^\beta$,
 \begin{equation}
 \frac{1}{2^n}\tr [ \sigma_i^\alpha(t) \sigma_j^\beta 
\sigma_i^\alpha(t) \sigma_j^\beta ] =  \begin{cases}
 1 \quad & 
[\sigma_j^\beta,\sigma_i^\alpha(t)] = 0; \\
  -1 \quad & 
\{\sigma_j^\beta,\sigma_i^\alpha(t)\} = 0.
 \end{cases}
 \end{equation}
 Taking $U$ randomly from the Clifford group, the operator $\sigma_i^\alpha(t)$ could be either one of all $4^n-1$ possible Pauli strings (the identity operator excluded) with equal probability. We can now compute $O_{\infty}$ by counting how many of the total $4^n-1$ possible Pauli strings commute with $\sigma_j^\beta$ and how many anti-commute:

\begin{enumerate}
 \item $[\sigma_j^\beta,\sigma_i^\alpha(t)] = 0$: $\sigma_j^\beta$ is a product of a Pauli matrix $\sigma_i^{\alpha}$ at site $i$ and identity operators at other sites. In order to $\sigma_j^\beta$ and $\sigma_i^\alpha(t)$ to commute, $\sigma_i^\alpha(t)$ must have either the identity operator or $\sigma_i^{\alpha}$ at site $i$ and an arbitrary operator at other sites. There are $2\cdot 4^{n-1}$ such Pauli strings, of which one must subtract the identity operator. To conclude, there are $2 \cdot 4^n - 1$ possible operators $\sigma_i^\alpha(t)$ out of the total $4^n-1$ which commute with $\sigma_j^\beta$.
 \item $\{\sigma_j^\beta,\sigma_i^\alpha(t)\} = 0$: in this case we can just calculate the number of operators $\sigma_i^\alpha(t)$ by subtracting the number of operators which commute with $\sigma_j\beta$ from the total $4^n-1$. We obtain $2 \cdot 4^{n-1}-1$.
\end{enumerate}

Putting this together we get $O_{\infty}$
\begin{equation}
 O_{\infty} = 1-\left( \frac{2 \cdot 4^n-1}{4^n-1} - \frac{2 
\cdot 4^n}{4^n-1} \right) = 1+\frac{1}{4^n-1} .
\end{equation}
For saturation value of $n$-point OTOC generalizations see Ref.~\cite{yoshida17}, for chaotic models Ref.~\cite{huang19}.

	\section{U(4) exact results}
	\label{app:U4_results}
	
	
	This appendix is dedicated to analytic solutions of OTOC time-dependence in a random quantum circuit with either BW OBC or BW PBC configurations and the choice of random two-site gate. Choosing a random two-site gate means that every elementary step is composed by a random gate uniformly drawn from the unitary group U(4) according to the Haar measure. All shall be calculated independently on the local Hilbert space dimension $q$ -- instead of qubits we shall now work with general qudits. The analytic solution when $q=2$ can be used as a non-trivial check of the exactness of our newly derived Markov chain. Namely, average dynamics for random two-site gate can be obtained by setting $u=0$, $v=-3/5$ in the transfer matrix from Eq.~\ref{eq:M_cd}.
	
	In order to analytically solve the time evolution of OTOC, we will heavily rely on a reduction of OTOC dynamics to an Ising-like partition function found in \cite{adam18}. Before we continue with solutions for BW OBC and BW PBC, we will give a brief overview of the reduction obtained in the previously mentioned paper. We won't give a detailed description of the reduction, but rather explain what one has to do in order to obtain the final result. 
	
\begin{figure}[h]
		\begin{center}
			\includegraphics[width=80mm]{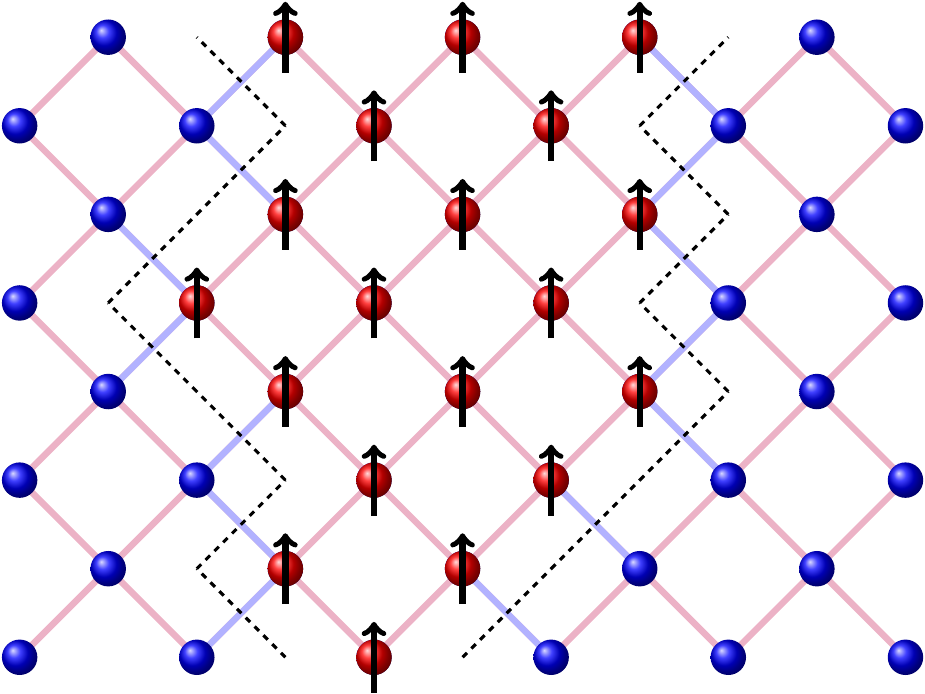}
			\caption{Graphical representation of the reduction of OTOC dynamics to a partition function of an Ising model. In the figure is represented one possible realization of the Ising model, which must be summed in the partition function. Red dots represent up-spins and blue dots represent down-spins. Interaction between spins are colored according to their weight ($1$ for same spins and $q/(q^2+1)$ for different spins). In the case shown here is represented a system with $12$ qudits up to time $\tau = 8$. At the lower boundary, we start with the up-spin located at the 5th (or equivalently 6th) qudit. The domain walls, describing the boundary between up and down-spins, must contain the qudit $j$ at time $\tau=8$ (for the example shown $j$ must be taken from the interval $\left[ 3, 8 \right]$ in order to the picture to represent a possible configuration). For the configuration shown in the appropriate term of the sum in Eq.~\ref{eq:OTOC_U4} one should take $``\mathrm{int.}=14"$ and $``\mathrm{width}=3"$.}
			\label{fig:partition_function.pdf}
		\end{center}
	\end{figure}

	The Ising-like model of which we will calculate the partition function is obtained by replacing all two-site elementary steps $U_{i,i+1}$ in the random quantum circuit with two-level spins ($s \in \{ \uparrow, \downarrow \}$). When dealing with a BW protocol, one obtains a grid of spins (see Fig.~\ref{fig:partition_function.pdf}). Contrary to the main text, here we measure time $\tau$ such that one unit corresponds to a row of the BW circuit. The choice of OTOC $O(i,j,\tau)$ reflects itself in the upper and lower boundary of the spin grid: at the lower boundary one must place an up-spin at position $(i,i+1)$ (for simplicity we shall assume that $i$ is odd), at the upper boundary one must place an up-spin at position $(j,j+1)$ or $(j,j-1)$, depending on time $\tau$ and parity of $j$. Moreover, boundary conditions dictate that all other spins at the lower boundary must be $\downarrow$. 
	
	Reference \cite{adam18} gives a detailed description on how to obtain such constraints, together with a derivation of the interaction laws between spins at different positions. Namely, spins interact with a three-body interaction. Take three spins, $(i,i+1)$ and $(i+2,i+3)$ at time $\tau$, and $(i+1,i+2)$ at time $\tau+1$. The weight of the interaction is $1$ if all spins are equally oriented, $q/(q^2+1)$ if the spins at time $\tau$ are different and $0$ otherwise. It is now clear that in order to obtain an up-spin at position $(j,j+1)$ at the upper boundary the spin grid must contain two domain walls, determining the boundary between up-spins and down-spins (see Fig.~\ref{fig:partition_function.pdf}). The authors of \cite{adam18} showed that OTOC are now recovered with the following equation
	
	\begin{equation}
		O(i,j,\tau) = \frac{q^2}{q^2-1} \sum_{\mathrm{ends}} (\frac{q}{q^2+1})^{\mathrm{int.}} (q^2)^{\mathrm{widht}} \# \mathrm{walls}.
		\label{eq:OTOC_U4}
	\end{equation}
	The first term in Eq.~\ref{eq:OTOC_U4} comes from the lower boundary condition and will be present in all configurations (PBC and OBC) that we study here. The sum runs over all possible ends of the domain walls on the upper boundary with width ``width". With ``int." we denoted the number of interactions between differently oriented spins. In the sum we must count all possible domain walls with given ``ends" ($\# \mathrm{walls}$). For a more graphical explanation of Eq.~\ref{eq:OTOC_U4} see Fig.~\ref{fig:partition_function.pdf}.
	
	In \cite{adam18}, OTOC were calculated for an infinite size system (infinite number of qudits). The authors obtained
	
	\begin{align}
		O(i,j,\tau) &=  \\
		 \zeta &g\left( \tau-1,\frac{\tau-\Delta j-1}{2},p \right) g\left( \tau-1,\frac{\tau+\Delta j-1}{2},p \right) \nonumber \\
		 +(1-\zeta) &g\left( \tau-1,\frac{\tau-\Delta j-3}{2},p \right) g\left( \tau-1,\frac{\tau+\Delta j-3}{2},p \right), \nonumber
	\label{eq:U4_inf}
	\end{align}
	with $\Delta j = |j-i|$ and
	\begin{align}
		&\zeta = \frac{q^4}{q^4-1} \\
		&p = \frac{1}{q^2+1} \\
		&g\left(n,a,p\right) = \sum_{k=0}^{a} \binom{n}{k} (1-p)^{n-k}p^k.
	\end{align}

	Infinite size systems can be thought as a BW PBC or BW OBC at early times, namely, all these systems share the same dynamics up to times $t \sim n$, where $n$ is the number of qudits (before the time evolved operator in $O(i,j,t)$ reaches the boundary). This means that by calculating the time derivative of $-2 \ln |O(j,\tau)-O_{\infty}|$ for $\tau \rightarrow \infty$ we would obtain the phantom eigenvalue for BW PBC circuits (when $n\gg 1$). Note that the prefactor $2$ comes from our time definition. In order to compute the limit, we will simplify Eq.~\ref{eq:U4_inf} by taking $\tau \gg j$ and $\tau \gg 1$
	
	\begin{equation}
		O(i,j,\tau) \approx g\left( \tau,\frac{\tau}{2},\frac{1}{q^2+1} \right)^2.
	\end{equation}
	In this case $\lim_{\tau \rightarrow \infty} g\left( \tau,\frac{\tau}{2},\frac{1}{q^2+1} \right) = 1$, so we replace $O_\infty$ with $1$. By taking the limit of $-2 \ln |g\left( \tau,\frac{\tau}{2},\frac{1}{q^2+1} \right)^2-1|$ we obtain our final result
	
	\begin{equation}
		\lim_{\tau \rightarrow \infty} -2 \ln |g\left( \tau,\frac{\tau}{2},\frac{1}{q^2+1} \right)^2-1| = 2 \ln \frac{1 + q^2}{2 q},
		\label{eq:OBC_slope}
	\end{equation}
	which gives the right result for $q=2$: $2 \ln{\frac{5}{4}}$ \cite{adam18,Frank18,AdamPRB19}.
	
\subsection{OTOC in a finite system with open boundary conditions}
	
	We shall now compute $O(1,j,\tau)$ for a finite size BW OBC circuit, where the propagated operator $\sigma_i^\alpha(t)$ is initially located near the boundary ($i=1$). Other choices of the initial location of $\sigma_i^\alpha(t)$ are less interesting, because $O(i,j,\tau)$ are identical to OTOC from Eq.~\ref{eq:U4_inf} for times $t \lesssim n$. Positioning $\sigma_i^\alpha(t)$ near the boundary thus gives new results for all times. Moreover, positioning $\sigma_i^\alpha(t)$ at the leftmost position guarantees that there exist only one domain wall. In this case we have to count only interactions between opposite spins at the right domain wall.
	
	In order to calculate $O(1,j,\tau)$ we must determine every quantity in Eq.~\ref{eq:OTOC_U4}. From the previous discussion we learned that there exists only one domain wall, which propagates in the system with increasing time. At time $\tau$ the up-spin domain must cover the $j$-th spin if we want the configuration to contribute to $O(1,j,\tau)$. Dealing with OBC circuits one can differentiate between two different up-spin domain scenarios: a) the domain is never wider than $n/2$ spins, b) the domain reaches width $n/2$ at time $\tau$, after this time there are no down-spins. This distinction between different domain scenarios will help us determine all the terms in Eq.~\ref{eq:OTOC_U4}: we shall separately count domain walls that reach width $n/2$ and domain wall that never reach width $n/2$. 
	
\begin{figure}[t]
	\begin{center}
		\includegraphics[width=80mm]{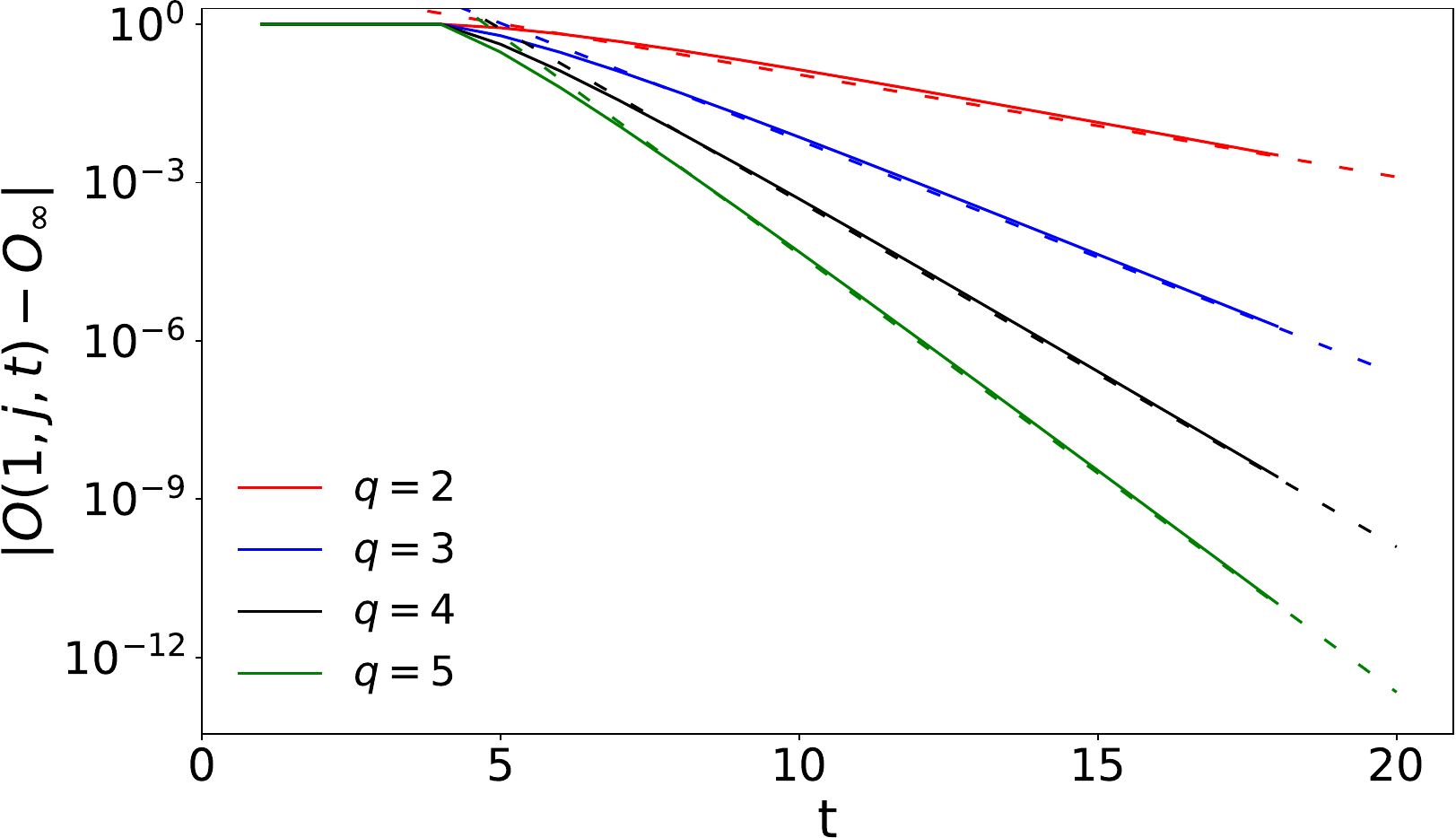}
		\caption{Time dependence of $O(1,j=10,t)$ for different choices of the local dimension $q$, $n=20$. With dashed lines are plotted lines with slope $2 \ln \frac{1 + q^2}{2 q}$ corresponding to the convergence rate of OTOC in an infinite system (see Eq.~\ref{eq:OBC_slope}). }
		\label{fig:OBC_q}
	\end{center}
\end{figure}

	The sum in Eq.~\ref{eq:OTOC_U4} can be separated as follows: 
	
	\begin{itemize}
		\item All configuration with up-spin domain of width $n/2$. In this case $(q^2)^\mathrm{width} = q^n$ and the number of interactions between opposite spins corresponds to the time $\tau_0$ when the up-spin domain spreads over $n$ qudits. The number of different spin configurations can be counted as the sum over all times of the number of domain walls that reach the rightmost position for the first time at time $\tau_0$ without reaching the leftmost position. The number of different paths $H(\tau,n)$ was obtained by subtracting any path that reach $n$ before time $\tau_0$ from the number of all possible paths
		
		\begin{align}
			H(\tau,n) = p(\tau,n) - \sum_{j=n/2}^{k/2-1} p(2j,n) \Delta_{\tau-2j}
		\end{align}
		where 
		\begin{equation}
			p(k,n) = \begin{cases}
						\frac{n}{k} \binom{k}{\frac{k-n}{2}} \quad &,\frac{k-n}{2} \mathrm{even} \\
						0 &,\mathrm{otherwise}
					\end{cases}
		\end{equation}
		counts the number of domain walls which reach point $n$ at time $k$ without ever crossing the point $0$ and where 
		
		\begin{align}
			\Delta_{2k} = -\sum_{N} \prod_{p} (-1)^p \Delta(2 r_p,n),
		\end{align}
		with $N=2^{k-1}$ denoting the number of compositions of $k$, where the order of different terms matters (for example compositions $(2,4)$ and $(4,2)$ of the integer $6$ are two different terms in the upper sum). The product runs over all $p$ terms $r_p$ of a composition, namely $k = \sum_p r_p$. The function $\Delta(k,j)$ is defined as
		
		\begin{equation}
			\Delta(k,j) = \binom{k}{\frac{k-n+j}{2}} - \binom{k}{\frac{k-n-j}{2}}
		\end{equation}
		and counts the number of different paths that start from $n$ and reach $j$ at time $k$ and never cross the position $0$.
		
		Configurations with width $n/2$ thus contribute with
		
		\begin{align}
			\frac{q^2}{q^4-1} q^n \sum_{\tau_0 = 1}^{\tau} (\frac{q}{q^2 + 1})^{\tau_0-1} H(\tau_0+1,n)	\label{eq:U4_OBC_1}
		\end{align}
		
		\item All configuration with width always smaller than $n/2$. In this case $(\frac{q}{q^2+1})^{\mathrm{int.}}=(\frac{q}{q^2+1})^{\tau-1}$. The number of different spin configurations will be computed as the sum over all domain wall endpoints (so that up-spins contain the qudit $j$) that do not reach $0$ minus all domain walls that at arbitrary times hit the right boundary $n$, that is $H(\tau_0,n)$. The factor $(q^2)^\mathrm{width}$ depends on the endpoint of the domain wall. Such configurations contribute to the partition function as

		\begin{align}
			&\frac{q^2}{q^4-1} (\frac{q}{q^2 + 1})^{\tau-1}  \sum_{u=j}^{n-1} q^u [ p(\tau+1,u) - \label{eq:U4_OBC_2} \\
			&\sum_{\tau_0=1}^{\tau} H(\tau_0,n) \Delta(\tau+1-\tau_0,u) ]. \nonumber
		\end{align}
		
	\end{itemize}
	A detailed explanation on how the numbers of paths were obtained will not be given here, because it deviates too much from the topic. OTOC $O(1,j,\tau)$ for BW OBC with random 2-site gates is equal to the sum of Eq.~\ref{eq:U4_OBC_1} and Eq.~\ref{eq:U4_OBC_2}. The obtained result was used to plot the time dependence of OTOC $O(j=10,t)$ for different values of $q$ for a system with $20$ qudits, see Fig.~\ref{fig:OBC_q}.	

\subsection{OTOC in a finite system with periodic boundary conditions}
	
	The case of BW PBC is more complicated. In this case our analytical result does not give any computational advantage over the Markov chain iteration method. We will see that the final results will be given by a recursion, which is time consuming during numerical evaluations. However, it is still useful to obtain an analytical result for completeness. As in the OBC case, here we also differentiate configuration that are wide $n/2$ and configurations that are never wider than $n/2$. 

	\begin{itemize}
	\item The number of domain walls with width $n/2$ at time ${\tau_c}$ will be computed recursively. Let $\hat{N}_{\tau_c}(u_0 \rightarrow u_1,v_0 \rightarrow v_1)$ denote the number of domain walls pairs, starting at $(u_0,v_0)$ and ending at $(u_1,v_1=u_1+n/2)$. All domain walls in $\hat{N}_{\tau_c}(u_0 \rightarrow u_1,v_0 \rightarrow v_1)$ must never be wider than $n/2$ at times $\tau'<{\tau_c}$. To shorten the notation, we shall always take $u_0 = 0$ and $v_1=1$, i.e., $i=1$, and write $\hat{N}_{\tau_c}(u_0 \rightarrow u_1, v_0 \rightarrow v_1)$ as $\hat{N}^{(\tau_c)}_{u_1}$. The quantity $\hat{N}^{(\tau_c)}_{u_1}$ can be computed recursively as
	
	\begin{align}
        \hat{N}^{(\tau_c)}_{u_1} &= N^{(\tau_c)}_{(u_0,v_0) \rightarrow (u_1,v_1)} - \label{eq:N_hat} \\
        \sum_{\tau_0 = n/2}^{{\tau_c}-1} \sum_{u'=0}^{\tau_0-n/2} & \hat{N}^{(\tau_0)}_{u'} N^{(\tau_c-\tau_0+1)}_{(u',u'+n/2) \rightarrow (u_1,v_1)}, \nonumber
    \end{align}
	where $N^{(\tau)}_{(u_0,v_0) \rightarrow (u_1,v_1)}$ counts the number of non-intersecting domain walls from $(u_0,v_0)$ to $(u_1,v_1)$, namely
	
	\begin{align}
        N^{(\tau)}_{(u_0,v_0) \rightarrow (u_1,v_1)} =&\\
        \binom{{\tau}-1}{u_1-u_0} \binom{{\tau}-1}{v_1-v_0}-&\binom{{\tau}-1}{v_1-u_0} \binom{{\tau}-1}{u_1-v_0}, \nonumber
	\end{align}
    where binomial coefficients $\binom{n}{k} = 0$ for $n<k$ or $k<0$. The recursion in Eq.~\ref{eq:N_hat} ends at $\tau_c=n/2$ by taking $\hat{N}^{(\tau_c)}_u = 1$.

	The contribution of domain walls of width $n/2$ follows immediately 
	
	\begin{equation}
		q^n \frac{q^2}{q^4-1} \sum_{{\tau_c}=n/2}^{\tau-1} \sum_{u=0}^{{\tau_c}-n/2}(\frac{q}{q^2 + 1})^{2\tau_c -2} \hat{N}^{(\tau_c)}_u \label{eq:U4_PBC_1}.
	\end{equation}
	
	\item Domain walls with width always smaller than $n/2$ will be computed with the help of $\hat{N}^{(\tau_c)}_{u_1}$. The number of desired paths will be calculated by subtracting domain walls that reach width $n/2$ from the total number of all possible paths
	
	\begin{align}
		\frac{q^2}{q^4-1} &(\frac{q}{q^2 + 1})^{2\tau - 2} \sum_{u=0}^{\tau-1} \sum_{v=v_b}^{u+n/2} q^{2(v-u)} \label{eq:U4_PBC_2} \\
        [ N^{(\tau)}_{(0,1) \rightarrow (u,v)} & - \sum_{\tau_0=n/2}^{\tau-1} \sum_{u'=0}^{\tau_0-n/2} \hat{N}^{(\tau_0)}_{u'} N^{(\tau-\tau_0+1)}_{(u',u'+n/2) \rightarrow (u,v)}] \nonumber
	\end{align}
	with $v_b = u+\mod(D-\mod(-\tau+2u+n/2,n),n)/2+1$ and $D = n/2-1+j$ (note that we always take $i=1$).
	
	\end{itemize}	

	The final result is obtained by summing Eq.~\ref{eq:U4_PBC_1} and Eq.~\ref{eq:U4_PBC_2}. Note that due to recursive terms, this analytical result does not give any substantial computational advantage over the Markov chain iteration method, however we do not claim that there is no way to count the number of domain walls in a simpler way.

	\section{Randomness}
	\label{app:randomness}
\begin{figure*}[t]
		\begin{center}
			\includegraphics[width=0.9\textwidth]{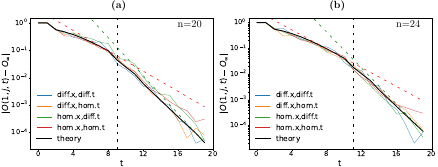}
			\caption{OTOC dynamics for a single realization of the random BW PBC circuit with dual-unitary gates $\az=0.5$, $j=4$, showing self-averaging. We show four different cases of selecting single-qubit random unitaries (see text) and theoretical averaged prediction based on our Markovian mapping (full black curve). Two dashed exponential functions are given by theoretical rates (red by Eq.(\ref{eq:lambda_PBC_S}, green by Eq.(\ref{eq:lambda_PBC_BW})). One can see that even though $n=24$ is not yet in the thermodynamic limit, at $t_c$ (vertical dashed line) the relaxation rate does change also for individual circuit realizations.}
			\label{fig:rand}
		\end{center}
	\end{figure*}

	In the main part of the paper we dealt with average OTOC dynamics (averaged over single-qubit random unitaries). Here we would like to see if the relaxation is similar without averaging, that is, for a single circuit realization. In order to explore how important is the choice of random one-site unitaries, we will study four different scenarios: 
	\begin{enumerate}
		\item When all one-site unitaries are independently drawn from the group U(2). In the following we shall denote this choice as diff.x,diff.t (different unitaries for every position and time). This choice was used in the main part of the paper;
		\item When the same one-site unitary is used at every time for the same position. Unitaries corresponding to different positions are drawn independently. In the following we shall denote this choice as diff.x,hom.t (different unitaries for every position but same unitaries for every time at fixed position);
		\item When the same one-site unitary is used for every position. At each time we generate a new independent one-site operator. In the following we shall denote this choice as hom.x,diff.t (different unitaries for every time but same unitaries for every position at fixed time);
		\item When at time $t=0$ we generate only one one-site random unitary and we repeat the same operator for all times and positions. In the following we shall denote this choice as hom.x,hom.t (same unitary for every position and time);
	\end{enumerate}
	Fig.~\ref{fig:rand} shows $|O(1,j=4,t)-O_{\infty}|$ for one realization of a BW PBC circuit with the gate $W$ having $\ax=1$, $\ay=1$ and $\az=0.5$. We can see that for all four cases there is a change in the slope around $t_c$ so we conjecture that in the thermodynamic limit the dynamics is self-averaging for each of the four scenarios, including the case where the random single-qubit unitary is the same at all qubits and at all times. This has interesting implications: first, explicit averaging over independent Haar random single-qubit unitaries in a random circuit is not necessary in order to observe phantoms, and second, it suggest that phantom eigenvalues, and with it a step-wise relaxation, could perhaps occur in other situations, not just in random circuits.
        \onecolumngrid
        
\begin{thebibliography}{99}

\bibitem{Hayden07} P.~Hayden and J.~Preskill, \tit{Black holes as mirrors: quantum information in random subsystems} J.~High Energ.~Phys. {\bf 2007}, 120 (2007).

\bibitem{Susskind08} Y.~Sekino and L.~Susskind, \tit{Fast scramblers} J.~High Energ.~Phys. {\bf 2008}, 065 (2008).

\bibitem{emerson03} J.~Emerson, Y.~S.~Weinstein, M.~Saraceno, S.~Lloyd, and D.~G.~Cory, \tit{Pseudo-random unitary operators for quantum information processing} Science {\bf 302}, 2098 (2003).

\bibitem{lashkari13} N. Lashkari, D. Stanford, M. Hastings, T. Osbornee, and P. Hayden, \tit{Towards the fast scrambling conjecture} J.~High Energ.~Phys. {\bf 2013}, 22 (2013). 

\bibitem{shenker14} S.~H.~Shenker and D.~Stanford, \tit{Multiple shocks} J.~High Energ.~Phys. {\bf 2014}, 46 (2014).

\bibitem{maldacena16} J. Maldacena, S. H. Shenker, and D. Stanford, \tit{A bound on chaos} J.~High Energ.~Phys. {\bf 2016}, 106 (2016).

\bibitem{roberts15} D. A. Roberts and D. Stanford, \tit{Diagnosing chaos using four-point functions in two-dimensional conformal field theory} Phys.~Rev.~Lett. {\bf 115}, 131603 (2015).

\bibitem{swingle17} D.~Chowdhury and B.~Swingle, \tit{Onset of many-body chaos in the O(N) model} Phys. Rev. D {\bf 96}, 065005 (2017).

\bibitem{moessner17} B. Dora and R. Moessner, \tit{Out-of-time-ordered density correlators in Luttinger liquids} Phys.~Rev.~Lett. {\bf 119}, 026802 (2017).

\bibitem{knap17} A. Bohrdt, C.B. Mendl, M. Endres, and M. Knap, \tit{Scrambling and thermalization in a diffusive quantum many-body system} New J.~Phys. {\bf 19}, 063001 (2017).

\bibitem{ueda17} N. Tsuji, P. Werner, and M. Ueda, \tit{Exact out-of-time-ordered correlation functions for an interacting lattice fermion model} Phys.~Rev.~A {\bf 95}, 011601(R) (2017).

\bibitem{smith19} A.~Smith, J.~Knolle, R.~Moessner, and D.L. Kovrizhin, \tit{Logarithmic spreading of out-of-time-ordered correlators without many-body localization} Phys.~Rev.~Lett. {\bf 123}, 086602 (2019).

\bibitem{nakamura19} S. Nakamura, E. Iyoda, T. Deguchi, and T. Sagawa, \tit{Universal scrambling in gapless quantum spin chains} Phys.~Rev.~B {\bf 99}, 224305 (2019).

\bibitem{xu19} S. Xu and B. Swingle, \tit{Locality, quantum fluctuations, and scrambling} Phys.~Rev.~X {\bf 9}, 031048 (2019).

\bibitem{lin18} C.-J. Lin and O.I. Motrunich, \tit{Out-of-time-ordered correlators in a quantum Ising chain} Phys.~Rev.~ B {\bf 97}, 144304 (2018).

\bibitem{oliveira} R.~Oliveira, O.~C.~O.~Dahlsten, and M.~B.~Plenio, \tit{Generic entanglement can be generated efficiently} Phys.~Rev.~Lett. {\bf 98}, 130502 (2007). 

\bibitem{PRA08} M.~\v Znidari\v c, \tit{Exact convergence times for generation of random bipartite entanglement} Phys.~Rev.~A {\bf 78}, 032324 (2008).

\bibitem{adam18} A.~Nahum, S.~Vijay, and J.~Haah, \tit{Operator spreading in random unitary circuits} Phys.~Rev.~X {\bf 8}, 021014 (2018).

\bibitem{Frank18} C.~W.~{von Keyserlingk}, T.~Rakovszky, F.~Pollmann, and S.~L.~Sondhi, \tit{Operator hydrodynamics, OTOC, and entanglement growth in systems without conservation laws} Phys.~Rev.~X {\bf 8}, 021013 (2018).

\bibitem{zhou19} T. Zhou and X. Chen, \tit{Operator dynamics in a Brownian quantum circuit} Phys.~Rev.~E {\bf 99}, 052212 (2019).

\bibitem{DU-LC} B.~Bertini, P.~Kos, and T.~Prosen, \tit{Exact correlation functions for dual-unitary lattice models in 1 + 1 dimensions} Phys.~Rev.~Lett {\bf 123}, 210601 (2019).

\bibitem{Bruno20} B.~Bertini and L.~Piroli, \tit{Scrambling in random unitary circuits: Exact results} Phys.~Rev.~B {\bf 102}, 064305 (2020).

\bibitem{lamacraft21} P.~W.~Claeys, J.~Herzog-Arbeitman, and A.~Lamacraft, \tit{Correlations and commuting transfer matrices in integrable unitary circuits} SciPost Phys. {\bf 12}, 007 (2022).

\bibitem{maximum_velocity} P.~W.~Claeys and A.~Lamacraft, \tit{Maximum velocity quantum circuits} Phys.~Rev.~Res. {\bf 2}, 033032 (2020).

\bibitem{kos21} P.~Kos, B.~Bertini, and T.~Prosen, \tit{Correlations in perturbed dual-unitary circuits: efficient path-integral formula} Phys.~Rev.~X {\bf 11}, 011022 (2021).

\bibitem{hashimoto20} K. Hashimoto, K.-B. Huh, K.-Y. Kim, and R.Watanabe, \tit{Exponential growth of out-of-time-order correlator without chaos: inverted harmonic oscillator} J.~High Energ.~Phys. {\bf 2020}, 68 (2020)

\bibitem{cao20} T.~Xu, T.~Scaffidi, and X.~Cao, \tit{Does scrambling equal chaos?} Phys.~Rev.~Lett. {\bf 124}, 140602 (2020).

\bibitem{santos20} S. Pilatowsky-Cameo, J. Chavez-Carlos, M. A. Bastarrachea-Magnani, P. Stransky, S. Lerma-Hernandez, L.F. Santos, and J.G. Hirsch, \tit{Positive quantum Lyapunov exponents in experimental systems with a regular classical limit} Phys.~Rev.~E {\bf 101}, 010202(R) (2020).

\bibitem{sarang18} S. Gopalakrishnan, D.A. Huse, V. Khemani, and R. Vasseur, \tit{Hydrodynamics of operator spreading and quasiparticle diffusion in interacting integrable systems} Phys.~Rev.~B {\bf 98}, 220303(R) (2018).

\bibitem{riddell21} J.~Riddell, W.~Kirkby, D.H.J. O’Dell, and E.S. Sorensen, \tit{Scaling at the OTOC wavefront: integrable versus chaotic models} arXiv:2111.01336 (2011).

\bibitem{khemani18} V.~Khemani, D.~A.~Huse, and A.~Nahum, \tit{Velocity-dependent Lyapunov exponents in many-body quantum, semiclassical, and classical chaos} Phys.~Rev.~B {\bf 98}, 144304 (2018).

\bibitem{saso17} I.~Kukuljan, S.~Grozdanov, and T.~Prosen, \tit{Weak quantum chaos} Phys.~Rev.~B {\bf 96}, 060301 (2017).

\bibitem{prejsnji_clanek} J.~Bensa, M.~\v Znidari\v c, \tit{Fastest local entanglement scrambler, multistage thermalization, and a non-Hermitian phantom} Phys.~Rev.~X {\bf 11}, 031019 (2021).

\bibitem{Mori20} T.~Mori and T.~Shirai, \tit{Resolving a discrepancy between Liouvillian gap and relaxation time in boundary-dissipated quantum many-body systems} Phys.~Rev.~Lett. {\bf 125}, 230604 (2020).

\bibitem{ueda21} T.~Haga, M.~Nakagawa, R.~Hamazaki, and M.~Ueda, \tit{Liouvillian skin effect: slowing down of relaxation processes without gap closing} Phys.~Rev.~Lett. {\bf 127}, 070402 (2021).

\bibitem{dekompozicija_1} N.~Khaneja, R.~Brockett, and S.~J.~Glaser, \tit{Time optimal control in spin systems} Phys. Rev. A {\bf 63}, 032308 (2001).

\bibitem{dekompozicija_2} B.~Kraus and J.~I.~Cirac, \tit{Optimal creation of entanglement using a two-qubit gate} Phys. Rev. A {\bf 63}, 062309 (2001).

\bibitem{dekompozicija_recept} M.~Blaauboer and R.~L.~de~Visser, \tit{An analytical decomposition protocol for optimal implementation of two-qubit entangling gates} J. Phys. A: Math. Theor. {\bf 41}, 395307 (2008).
	
\bibitem{metoda_redukcija} W.-T.~Kuo, A.~A.~Akhtar, D.~P.~Arovas, and Y.~Z.~You, \tit{Markovian entanglement dynamics under locally scrambled quantum evolution} Phys. Rev. B {\bf 101}, 224202 (2020). 
   
\bibitem{oliveira07} O.~C.~O.~Dahlsten, R.~Oliveira, and M.~B.~Plenio, \tit{The emergence of typical entanglement in two-party random processes} J.~Phys.~A {\bf 40}, 8081 (2007).

\bibitem{foot0}  One could ask whether the time evolution of $\Phi$ constructed from coefficients $c_\mathbf{s}$ could be in general different from the time evolution of $\Phi$ obtained from $a_\mathbf{s}$, because the former coefficients are obtained from a density matrix $\rho$, the latter from a Pauli matrix. However, one can check that the density matrix $\rho$ in the Markov chain derivation from \cite{metoda_redukcija} can be substituted with any Hermitian operator.  


\bibitem{ballistic} D.~A.~Roberts, D.~Stanford, and L.~Susskind, \tit{Localized shocks} J. High Energy Phys. {\bf 2015}, 51 (2015).

\bibitem{vB_1} I.~L.~Aleiner, L.~Faoro, and L.~B.~Ioffe, \tit{Microscopic model of quantum butterfly effect: Out-of-time-order correlators and traveling combustion waves} Ann. Phys. {\bf 375}, 378 (2016).

\bibitem{vB_2} A.~A.~Patel, D.~Chowdhury, S.~Sachdev, and B.~Swingle, \tit{Quantum butterfly effect in weakly interacting diffusive metals} Phys.~Rev.~X {\bf 7}, 031047 (2017).

\bibitem{LiebRobinson} E.~H.~Lieb and D.~W.~Robinson, \tit{The finite group velocity of quantum spin systems} Commun. Math. Phys. {\bf 23}, 251 (1972).

\bibitem{yoshida17} D.~A.~Roberts and B.~Yoshida, \tit{Chaos and complexity by design} J.~High Energ.~Phys. {\bf 2017}, 121 (2017).

\bibitem{Znidaric_2007} M.~\v Znidari\v c, \tit{Optimal two-qubit gate for generation of random bipartite entanglement} Phys.~Rev.~A {\bf 76}, 012318 (2007).

\bibitem{foot1}
		Based on numerical calculations we conjecture that $|\lambda_2|$ for S PBC and arbitrary dual-unitary gates in the TDL is given by
		
		\begin{equation}
			|\lambda_2|_{\mathrm{S-PBC}} = \frac{1}{3}(2-\cos(\az \pi)),
			\label{eq:lambda_PBC_S}
		\end{equation}
		and for BW PBC and dual-unitary by
		\begin{equation}
			|\lambda_2|_{\mathrm{BW-PBC}} = |\lambda_2|_{\mathrm{S-PBC}}^2 = \frac{1}{9}(2-\cos(\az \pi))^2.
\label{eq:lambda_PBC_BW}
		\end{equation}
		These equations will be used to check whether phantoms from S PBC are present in dual-unitary random quantum circuits.

\bibitem{sarang21}  S. Gopalakrishnan and M.J. Gullans, \tit{Entanglement and purification transitions in non-Hermitian quantum mechanics} Phys.~Rev.~Lett. {\bf 126}, 170503 (2021).

\bibitem{mori21} T.~Mori, \tit{Metastability associated with many-body explosion of eigenmode expansion coefficients} Phys.~Rev.~Res. {\bf 3}, 043137 (2021).

\bibitem{huang19} Y.~Huang, F.G.S.L. Brand\~ ao, and Y.-L. Zhang, \tit{Finite-size scaling of out-of-time-ordered correlators at late times} Phys.~Rev.~Lett. {\bf 123}, 010601 (2019).

\bibitem{AdamPRB19} T.~Zhou and A.~Nahum, \tit{Emergent statistical mechanics of entanglement in random unitary circuits} Phys.~Rev.~B {\bf 99}, 174205 (2019).

                

	\end{thebibliography}
\end{document}